\documentclass[a4paper,11pt]{article}
\usepackage{jheppub} 
\usepackage{lineno}
\usepackage{slashed}
\usepackage{multirow,multicol}
\usepackage{amsmath}
\usepackage{physics}
\usepackage{setspace}

\renewcommand{\O}{\mathcal{O}}
\renewcommand{\op}[3]{\O^{#2,#3}_{#1}}
\newcommand{\hc}{\mathrm{h.c.}}
\newcommand{\calo}{\mathcal{O}}
\newcommand{\lra}[1]{\langle#1 \rangle}
\newcommand{\dl}[1]{f_\chi\exp\left(#1\frac{\chi}{f_\chi}\right)}
\newcommand{\lrd}{\overleftrightarrow{D}}
\newcommand{\lqcd}{\Lambda_{\text{QCD}}}

\arxivnumber{2601.16534} 

\title{\boldmath Effective Field Theory Description of Light Dilaton}

\author{
  Qing-Hong Cao$^{1,2,3}$,
  Jian-Nan Ding$^{3}$,
  Bing-Hui Ge$^{4}$,
  Hao Sun$^{5,6,7}$, 
  Jiang-Hao Yu$^{6,7,8,9}$
}

\affiliation{\vspace{2mm} $^1$
Department of Physics and State Key Laboratory of Nuclear Physics and Technology, Peking University, Beijing 100871, China \\
$^2$School of Physics, Zhengzhou University, Zhengzhou 450001, China\\
$^3$Center for High Energy Physics, Peking University, Beijing 100871, China \\
$^4$School of Physics, Peking University, Beijing 100871, China\\
$^5$Institute of High Energy Physics, Chinese Academy of Sciences, Beijing 100049, China\\
$^6$Institute of Theoretical Physics, Chinese Academy of Sciences,   Beijing 100190, P. R. China\\
$^7$School of Physical Sciences, University of Chinese Academy of Sciences,   Beijing 100049, P.R. China\\
$^8$School of Fundamental Physics and Mathematical Sciences, Hangzhou Institute for Advanced Study, UCAS, Hangzhou 310024, China\\
$^9$International Centre for Theoretical Physics Asia-Pacific, Beijing/Hangzhou, China \\
}

\emailAdd{qinghongcao@pku.edu.cn}
\emailAdd{dingjn23@pku.edu.cn}
\emailAdd{gebinghui@stu.pku.edu.cn}
\emailAdd{hsun@ihep.ac.cn}
\emailAdd{jhyu@itp.ac.cn}

\abstract{Dilatons, the CP-even pseudo-Nambu-Goldstone bosons arising from spontaneous scale symmetry breaking, offer a compelling alternative to axion-like particles (ALPs) yet lack a comprehensive low-energy framework. We address this by constructing a systematic effective field theory (EFT) for the dilaton based on a manifestly scale-invariant regularization scheme. This approach derives universal linear couplings to the trace anomaly while preserving consistent renormalization group evolution. We establish a hierarchical EFT tower connecting the ultraviolet conformal sector to the infrared, encompassing the dilaton-extended SMEFT, low-energy EFT up to dimension-7, and a chiral Lagrangian describing meson and baryon interactions. We perform a comprehensive phenomenological analysis across two distinct mass regimes, where the dilaton manifests as either a conventional particle or a wave-like particle. For MeV-scale dilatons behaving as conventional particles, we obtain constraints from LHC production, semi-invisible $B$- and $K$-meson decays, and supernova cooling. For ultralight dilatons acting as dark matter, we project sensitivities for atomic clocks and atom interferometers. This unified EFT framework would pave the way for extended phenomenological studies across the full mass spectrum of the light dilaton. }

\begin{document}
\maketitle
\flushbottom

\section{Introduction}
\label{sec:intro}

In the absence of direct evidence for new physics at the electroweak scale, interest in light scalar and pseudo-scalar bosons has grown significantly in recent years. Such states emerge naturally in various extensions of the Standard Model (SM) as pseudo-Nambu-Goldstone bosons (pNGBs) resulting from the spontaneous symmetry breaking. Their existence leads to diverse, testable phenomenological signatures, motivating extensive theoretical and experimental efforts.

Axions and generic axion-like particles (ALPs) are the most prominent candidates for light pseudo-scalars beyond the SM. They arise from the spontaneous breaking of approximate global $U(1)$ symmetries. The QCD axion was originally proposed within the Peccei-Quinn mechanism to resolve the strong CP problem in quantum chromodynamics (QCD)~\cite{Kim:2008hd,DiLuzio:2020wdo}. However, QCD axions with masses between~1~MeV and tens of GeV are strongly disfavored~\cite{Peccei:2006as,Turner:1989vc} due to the strict relation between their mass and decay constant. Consequently, attention has shifted towards generic ALPs, which are free from this constraint. Their low-energy effective field theories (EFTs) are well developed~\cite{Brivio:2017ije,Galda:2021hbr,Bauer:2020jbp}, and the corresponding renormalization group equations have been thoroughly studied~\cite{Chala:2020wvs,Bonilla:2021ufe}, providing a solid foundation for phenomenological analyses across a broad range of energy scales~\cite{Graham:2015ouw,Irastorza:2018dyq,Ertas:2020xcc}. Particular focus has been placed on the ultralight regime~\cite{Abel:2017rtm,ADMX:2018gho}, with masses $\lesssim$~eV, where ALPs are viable dark matter (DM) candidates. 

Complementing the search for pseudo-scalar ALPs, scalar dilatons represent another compelling class of light bosons. 
As pNGBs associated with the spontaneous breaking of approximate scale (or conformal) invariance~\cite{Salam:1969bwb,Isham:1970gz,Isham:1971dv,Ellis:1970yd,Ellis:1971sa}, dilatons are ubiquitous in beyond the Standard Model (BSM) constructions.
Notably, lattice simulations of confining gauge theories near the conformal window imply the existence of a light scalar field~\cite{Appelquist:2016viq,LatKMI:2016xxi,LatKMI:2014xoh,Fodor:2016pls,Fodor:2012ty}, which is interpreted as a potential dilaton, and has been discussed in Refs.~\cite{Appelquist:2017wcg,Appelquist:2017vyy,Appelquist:2019lgk,Appelquist:2022mjb,Appelquist:2025tol}. The resulting EFT for the dilaton offers a robust framework for realistic composite Higgs models~\cite{Appelquist:2020bqj,Appelquist:2022qgl} and can accommodate forbidden dark matter candidates~\cite{Appelquist:2024koa}.
While the dilaton's static properties and couplings to SM fields have been established~\cite{Goldberger:2007zk,Appelquist:2010gy,Vecchi:2010gj,PhysRevD.87.115006}, the renormalization group (RG) evolution of these interactions remains comparatively understudied. Consequently, despite a wide array of experimental searches in atomic, molecular, and gravitational systems~\cite{Arvanitaki:2014faa, VanTilburg:2015oza, Hees:2016gop, Berge:2017ovy, Hees:2018fpg, Kennedy:2020bac, Filzinger:2023zrs, Geraci:2016fva, Arvanitaki:2016fyj, Badurina:2019hst, Buchmueller:2023nll}, a continuous EFT description that consistently connects the UV symmetry-breaking scale to low-energy IR phenomena is still lacking.

\begin{figure}[htbp]
    \centering
    \includegraphics[width=\linewidth]{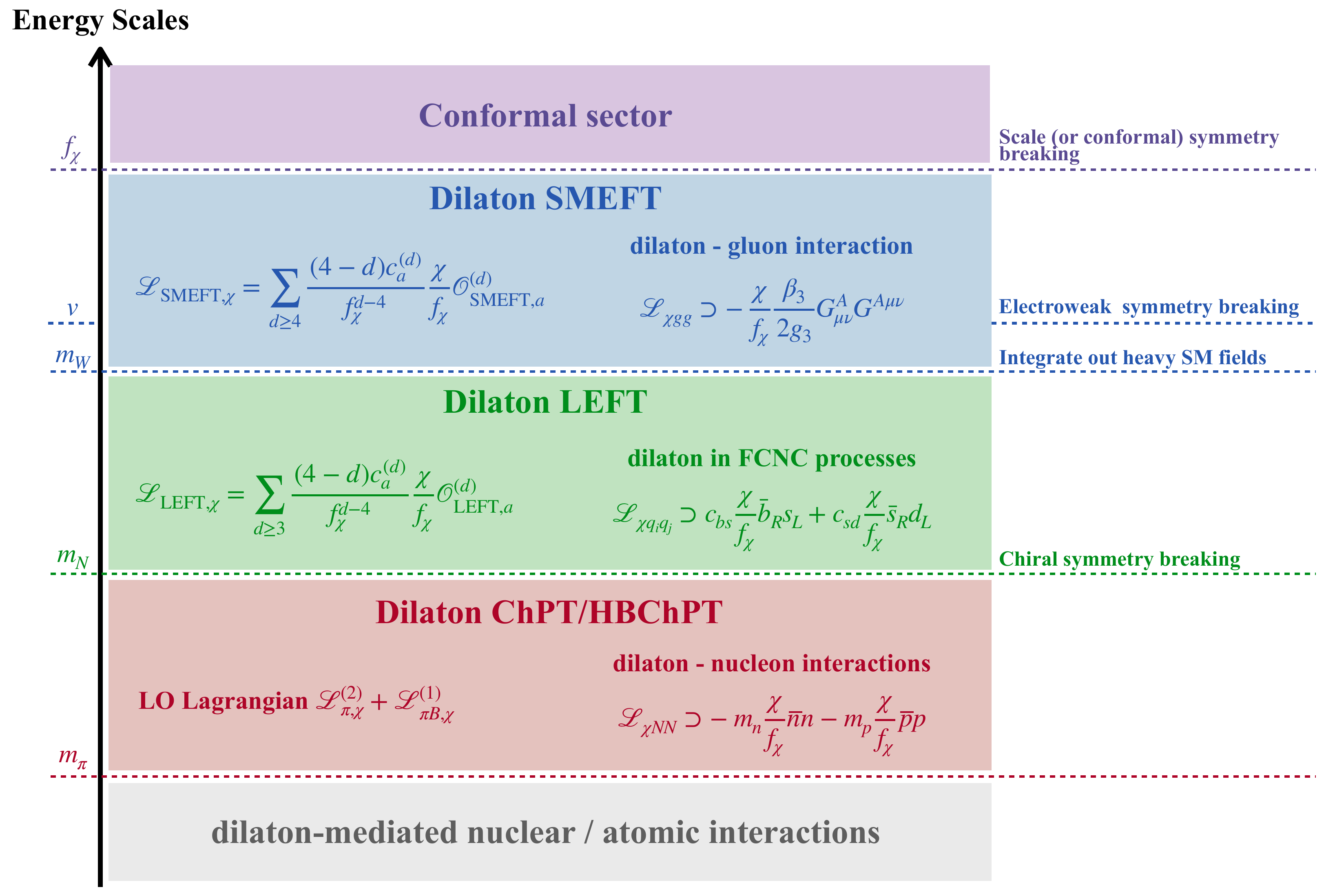}
    \caption{Schematic hierarchy of energy scales associated with the dilaton EFTs constructed in this work. The figure also shows the dilaton couplings to gauge bosons, quarks, and nucleons, which serve as inputs for the phenomenological analyses. These interactions can be probed via LHC searches, flavor-changing neutral currents (FCNCs), supernova cooling, and---assuming the dilaton accounts for the observed DM abundance---precision measurements of time-varying fundamental constants.}
    \label{fig:enegyscales}
\end{figure}

In this work, we address this gap by establishing a consistent particle-physics\footnote{Prior studies formulated in the context of general relativity have introduced the dilaton as a scalar field non-minimally coupled to the metric, with interactions invariant under Weyl transformations~\cite{Damour:2010rp,Nitti:2012ev}.} framework for the dilaton, originating from an ultraviolet (UV) conformal sector. While the scale of symmetry breaking lies significantly above the electroweak scale, the dilaton manifests as a light degree of freedom in the infrared (IR). This hierarchy necessitates the construction of a sequence of EFTs to rigorously bridge the disparate energy scales. We begin by treating the SM as the effective theory of the broken phase of a quantum scale-invariant UV completion, and subsequently extend the formalism to include higher-dimensional operators. We then build a tower of dilaton EFTs matched at successive energy scales. This hierarchical structure facilitates a model-independent description of dilaton interactions across different regimes, as illustrated in Fig.\,\ref{fig:enegyscales}.

Analogous to searches for ALPs, the detection strategy for the dilaton depends on its mass. A key distinction arises from the dilaton's scalar (CP-even) nature, which contrasts with the pseudo-scalar (CP-odd) couplings characteristic of ALPs. In this study, we focus on two representative mass regimes to derive complementary bounds on the dilaton parameter space. For masses near the MeV scale, the dilaton behaves as a particle-like state with interactions governed by the trace anomaly. In this regime, we probe the dilaton through signatures at the Large Hadron Collider (LHC), rare meson decays, and cosmological observations. Conversely, in the sub-eV regime, we treat the dilaton as a coherent wave-like field that induces oscillations in fundamental constants, detectable via precision measurements in atomic systems. While intermediate masses offer rich phenomenology, such as constraints from stellar cooling and violations of the Weak Equivalence Principle (WEP), we reserve these topics for future investigations.  

This paper is organized as follows: In Section~\ref{sec:framework}, we develop the theoretical framework for the dilaton, comparing two distinct construction approaches---the conformal compensator method and manifestly scale-invariant regularization---and verifying its universal coupling to the trace anomaly. In Section~\ref{sec:DilatonEFTs}, we derive its leading interactions with SM particles and extend the framework to higher-dimensional operators. The effective Lagrangian for the low-energy effective field theory (LEFT) extended by the dilaton is presented in Section~\ref{sec:LEFT} and subsequently matched onto a chiral perturbation theory ($\chi$PT) Lagrangian that includes the dilaton in Section~\ref{sec:ChEFT}. In Section~\ref{sec:pheno}, we derive complementary phenomenological constraints on the dilaton parameter space. Finally, we draw conclusions in Section~\ref{sec:summary}.

\section{Theoretical Framework}
\label{sec:framework}
In this section, we present the theoretical framework to construct an effective field theory for a light dilaton. When the conformal group $SO(2,4)$ spontaneously breaks down to the Poincaré group, although there are five broken generators associated with dilatations and special conformal transformations, only one physical Goldstone boson, the dilaton $\chi$, is required to restore conformal invariance in the effective Lagrangian~\cite{Low:2001bw}. At leading order in the dilaton field, its interactions with the visible-sector fields are usually given by a single term~\cite{Goldberger:2007zk}
\begin{equation}
    \mathcal{L}_\text{int}=-\frac{\chi}{f_\chi} T^\mu{}_\mu,
\end{equation}
where $\chi,f_\chi$ refer to the physical dilaton field and the dilaton VEV $f_\chi\equiv\langle \Phi\rangle$ respectively and $T^\mu_{\ \mu}$ is the trace of the improved energy-momentum tensor. Consequently, the linear dilaton couplings are controlled by the new parameter, the dilaton VEV. This universal structure can be derived in two physically equivalent ways: the conformal compensator method and the scale-invariant regularization using a dynamical subtraction scale $\mu(\Phi)$. Moreover, the renormalization-group evolution of all couplings in the dilaton effective theory coincides, at the one-loop level, with the running of the conventional theory without the dilaton once the renormalization scale is identified with the dilaton VEV in the broken phase. The following subsections present these results in detail.

\subsection{The trace of the energy-momentum tensor}
\label{sec:tensor-trace}
In this section, we review the structure of the energy-momentum tensor and its trace in a general field theory. Consider a theory described by a Lagrangian $\mathcal{L}(x)=\mathcal{L}(\{\phi(x),\partial_\mu\phi(x)\})$. 
Under scale transformation $x^\mu\rightarrow e^\lambda x^\mu$, a field $\phi$ of scaling dimension $d_\phi$ transforms as $\phi(x)\rightarrow e^{d_\phi\lambda}\phi(e^\lambda x)$. The associated Noether current in $d$ dimensions is
\begin{equation}
    J^\mu=\frac{\partial\mathcal{L}}{\partial(\partial_\mu\phi_\alpha)}(d_\phi+x^\mu\partial_\mu)\phi-x^\mu\mathcal{L}=x_\nu T^{\mu\nu},
\end{equation}
where $T^{\mu\nu}$ is the symmetric, conserved energy-momentum tensor~\cite{Callan:1970ze}. Its trace then equals the divergence of the scale current:
\begin{equation}
    T^\mu{}_\mu=\partial_{\mu} J^{\mu}=\frac{\partial \mathcal{L}}{\partial \phi} d_\phi \phi+\frac{\partial \mathcal{L}}{\partial \partial_{\mu} \phi}(d_\phi+1) \partial_{\mu} \phi-d \mathcal{L}.
    \label{equ:tensortrace}
\end{equation}
This relation implies that exact scale invariance requires a vanishing trace $T^\mu{}_\mu=0$. As an illustration, consider the QCD Lagrangian in $d=4-2\epsilon$ dimensions:
\begin{equation}
    \mathcal{L}_\mathrm{QCD}=\bar{\psi}(i \slashed{D}-m) \psi-\frac{1}{4g^2} G^a_{\mu\nu} G^{a\mu\nu},
    \label{equ:LQCD}
\end{equation}
where the field strength tensor $G^a_{\mu\nu}$ and covariant derivative $D_\mu$ are defined without explicit dependence on the gauge coupling $g$. Applying Eq.\,\eqref{equ:tensortrace}, the corresponding trace of the energy-momentum tensor is 
\begin{equation}
    T^\mu{}_\mu=m\bar{\psi}\psi-\frac{2\epsilon}{4g^2}G^a_{\mu\nu} G^{a\mu\nu}.
    \label{equ:tensortraceQCD}
\end{equation}
Renormalization of the composite operator $G^2\equiv G^a_{\mu\nu} G^{a\mu\nu}$ in the modified minimal
subtraction ($\overline{\mathrm{MS}}$) scheme gives the identity~\cite{Tarrach:1981bi}
\begin{equation}
    -2\epsilon\frac{G^2}{4}=\frac{\beta(g)}{2g}G^2+\gamma_m m\bar{\psi}\psi,
\end{equation}
where $\beta(g)$ is the QCD beta-function and $\gamma_m$ is the quark mass anomalous dimension. Inserting this relation and taking the $\epsilon\rightarrow0$ limit, one recovers the well-known QCD trace anomaly:
\begin{equation}
    T^\mu{}_\mu=(1+\gamma_m)m\bar{\psi}\psi+\frac{\beta}{2g^3}G^2.
\end{equation}

The above expression reveals that scale invariance is broken both classically by dimensional parameters (quark masses) and by the running of couplings at the quantum level. 
From this heuristic perspective of understanding the effects of scale transformations on the theory, a practical form of $T^\mu{}_\mu$ without the dimensional regulator $\epsilon$ can be derived by writing the Lagrangian in terms of the anomalous couplings and operators
\begin{equation}
    \mathcal{L}=\sum_i g_i(\mu)\mathcal{O}_i(x),
    \label{equ:generalL}
\end{equation}
where the coupling $g$ depends on the renormalization scale $\mu$ and the operator $\mathcal{O}_i$ has scaling dimension $d_i$. 
The scale transformation $x\rightarrow e^{\lambda}x$ is equivalent to shifting the renormalization scale as $\mu\rightarrow e^{\lambda} \mu$ while the operators transform as $\mathcal{O}_i(x)\rightarrow e^{\lambda d_i}\mathcal{O}_i(e^\lambda x)$. 
The variation of the Lagrangian under this combined transformation directly yields~\cite{Goldberger:2007zk}
\begin{equation}
T^\mu{}_\mu=\sum_i\beta(g_i)\mathcal{O}_i(x)+\sum_i(d_i-4)g_i(\mu)\mathcal{O}_i(x),
    \label{equ:tensortrace-practical}
\end{equation}
where $\beta(g_i)=\mu \frac{\partial g_i}{\partial\mu}$. This form manifestly separates explicit breaking by dimensional parameters from quantum anomalous breaking via running couplings.

\subsection{Conformal compensator method}
\label{sec:derivation1}
 
The dilaton is introduced through the conformal compensator field~\cite{Goldberger:2007zk} 
\begin{equation}
    \Phi=f_\chi e^{\chi/f_\chi},
\end{equation}
where $f_\chi=\langle\Phi\rangle$ is the order parameter for spontaneous conformal symmetry breaking, determined by the underlying conformal sector. Under scale transformations, $\chi$ transforms non-linearly as $\chi(x)\rightarrow\chi(e^\lambda x)+f_\chi\lambda$, while the compensator transforms linearly as 
\begin{equation}
    \Phi(x)\rightarrow e^{\lambda}{\Phi}(e^\lambda x).
\end{equation}

To make the Lagrangian in Eq.\,\eqref{equ:generalL} manifestly invariant, the dimensional couplings are treated as spurions by assigning them a fictitious scaling dimension $4-d_i$ and the anomalous breaking via running couplings is incorporated via renormalization scale's dependence on the dilaton field:
\begin{equation}
    g_i(\mu)\rightarrow g_i\left(\mu \frac{f_\chi}{\Phi}\right)\left(\frac{\Phi}{f_\chi}\right)^{4-d_i}.
\end{equation}
Expanding this scale-invariant Lagrangian to linear order in the dilaton field, we obtain
\begin{equation}
\begin{aligned}
     \mathcal{L}_\text{eff}=\sum_ig_i(\mu)O_i(x)-\sum_i\left[\beta(g_i)O_i(x)+\left(d_i-4\right)g_i(\mu)O_i(x)
    \right]\frac{\chi}{f_\chi},
\end{aligned}
\end{equation}
where the first term is the original Lagrangian $\mathcal{L}$, and the second term precisely gives the universal dilaton coupling to the trace of the energy-momentum tensor, $-\frac{\chi}{f_\chi}T^{\mu}_{\ \mu}$, with $T^{\mu}_{\ \mu}$ given in Eq.\,\eqref{equ:tensortrace-practical}. 

This construction ensures that low-energy dilaton interactions are dictated solely by the pattern of spontaneous scale (conformal) symmetry breaking, independent of the microscopic origin of the breaking sector.

\subsection{Manifestly scale-invariant regularization}
\label{sec:derivation2}
While the conformal compensator method offers a powerful and symmetry-driven framework for constructing the low-energy effective theory of the dilaton, it does not automatically provide a regularization and renormalization scheme that preserves scale invariance at the quantum level. A complementary approach, particularly well-suited for perturbative quantum computations in such effective theories, is the adoption of a manifestly scale-invariant regularization~\cite{Tamarit:2013vda,Ghilencea:2015mza}. In this section, we illustrate this method using QCD as a toy model. 

Assuming a UV-conformal sector beyond QCD, our starting point is the scale-invariant extension of the QCD Lagrangian in $d=4-2\epsilon$ dimensions, where the dilaton field $\Phi$ is introduced via analytical continuation:
\begin{equation}
    \mathcal{L}_\mathrm{inv}^{(d)}=\frac{1}{2}\partial_\mu\Phi\partial^\mu\Phi-\frac{\mu(\Phi)^{-2\epsilon}}{4g^2}G^2+\bar{\psi}i \slashed{D}\psi-y\mu(\Phi)^\epsilon\Phi\bar{\psi}\psi,
\end{equation}
where $y$ quantifies the tree-level Yukawa interaction between $\Phi$ and quarks. As in standard dimensional regularization, the couplings are rescaled by a renormalization scale $\mu$ to remain dimensionless. However, to preserve manifest scale invariance, we promote a constant scale $\mu$ to a field-dependent function 
\begin{equation}
    \mu(\Phi)=z\Phi^{\frac{1}{1-\epsilon}}
    \label{equ:mu(dilaton)},
\end{equation} 
where $z$ is an arbitrary dimensionless parameter. The exponent in Eq.\,\eqref{equ:mu(dilaton)} ensures that $\mu(\Phi)$ has mass dimension $1$.

After spontaneous scale symmetry breaking, the dilaton field $\Phi$ can be expanded around its VEV $f_\chi=\langle\Phi\rangle$, which represents the scale of new physics:
\begin{equation}
    \Phi=f_\chi+\chi,
\end{equation}
where the fluctuation field $\chi$ denotes the physical dilaton particle. The scale function $\mu(\Phi)$ now generates the usual renormalization scale $\mu_0\equiv \mu(f_\chi)$. For small fluctuations $\chi\ll f_\chi$, the effective Lagrangian reduces to 
\begin{equation}
\begin{aligned}
    \mathcal{L}^{(d)}=-\mu_0^{-2\epsilon}\frac{1}{4g^2}G^2 +\bar{\psi}i \slashed{D}\psi - y\mu_0^\epsilon f_\chi\bar{\psi}\psi-\frac{\chi}{f_\chi}\left[-\frac{\mu_0^{-2\epsilon}}{4g^2}2\epsilon G^2+y\mu_0^\epsilon f_\chi\bar{\psi}{\psi}\right],
    \label{equ:LdilatonQCDd}
\end{aligned}
\end{equation}
where higher-order terms in $\chi,\epsilon$ are neglected. The above Lagrangian corresponds to a renormalized theory at the scale $\mu_0$ and has two distinct components. The first three terms reproduce the standard QCD Lagrangian in Eq.\,\eqref{equ:LQCD} with a fermion mass $m=yf_\chi$ emerging dynamically. The dilaton interaction term is precisely proportional to the trace of the energy-momentum tensor $T^\mu{}_\mu$ given in Eq.\,\eqref{equ:tensortraceQCD}. This identification simplifies the Lagrangian in the limit $\epsilon\rightarrow0$
\begin{equation}
    \mathcal{L}_\text{eff}=-\frac{1}{4g^2}G^2+\bar{\psi}i \slashed{D}\psi - m\bar{\psi}\psi-\frac{\chi}{f_\chi} T^\mu{}_\mu.
    \label{equ:LdilatonQCD4}
\end{equation}
This structure is consistent with the expectations from the conformal compensator method. 

In this way, the standard QCD theory is embedded into a quantum scale-invariant framework, naturally generating linear dilaton couplings to the trace of the energy-momentum tensor. This approach preserves scale invariance manifestly during intermediate calculations, making it ideal for perturbative studies of dilaton effective theories. 

\subsection{Running couplings}
\label{sec:RGE}
In this section, we demonstrate that a quantum theory with spontaneously broken scale invariance, realized via the dilaton, can reproduce the running couplings of conventional theories with explicit dimensional parameters. Perturbative calculations are well-defined only in the broken phase, where the dilaton acquires a VEV. Building upon the effective scale-invariant extension of QCD involving a dilaton obtained in the previous section, we analyze the scaling properties of renormalized Green's functions in the broken phase, closely following the approach of Ref.~\cite{Tamarit:2013vda}. 

To illustrate the mechanism explicitly, we first recall the standard treatment in ordinary QCD, where the running of coupling $g$ and fermion mass $m$ with the renormalization scale $\mu$ is governed by the beta functions $\beta_g,\beta_m$. These are encoded in the scaling properties of the renormalized connected n-point Green's function. The Ward identity corresponding to the scale transformations in momentum space is given by 
\begin{equation}
    \left(\sum_{\phi, k} p_{k}^{\phi} \frac{\partial}{\partial p_{k}^{\phi}}+m\frac{\partial}{\partial m}+\mu \frac{\partial}{\partial \mu}+\sum_{\phi}\left(c_{\phi} n_{\phi}\right)-4\right) G^{\left(n_{\phi}\right)}\left(p_{k}^{\phi}\right)=0,
    \label{equ:QCD-ward}
\end{equation}
where $\phi=(G_\mu,\psi,\bar{\psi})$ labels the gluon, quark and anti-quark fields, $n_\phi=(n_G,n_\psi,n_{\bar{\psi}})$ denote the numbers of external legs of each type, $c_\phi=(3,\frac{5}{2},\frac{5}{2})$, and $p_k^\phi$ is the momentum for the $k$-th leg for the field $\phi$. This identity ensures the correct engineering dimension $4-\sum c_\phi n_\phi$ of the Green's function, accounting for the implicit $\mu$ dependence introduced by renormalization. The $\mu$ dependence is further constrained by the Callan-Symanzik equation, which expresses the independence of physical observables from the arbitrary renormalization scale $\mu$:
\begin{equation}
    \left(\mu \frac{\partial}{\partial \mu}+\beta_{g} \frac{\partial}{\partial g}+\beta_{m} \frac{\partial}{\partial m}+\sum_\phi n_\phi \gamma_\phi\right) G^{(n_\phi)}\left(p_{k}^\phi\right)=0,
    \label{equ:QCD-CS}
\end{equation}
with the beta functions and anomalous dimensions defined as
\begin{equation}
    \beta_g=\mu\frac{\partial g}{\partial\mu},\quad \beta_m=\mu\frac{\partial m}{\partial\mu},\quad \gamma_\phi=\frac{\mu}{2Z_\phi}\frac{\partial Z_\phi}{\partial\mu},
\end{equation} 
where $Z_\phi$ are the wave function renormalization constants. Combining Eq.\,\eqref{equ:QCD-ward} and Eq.\,\eqref{equ:QCD-CS} leads to the differential equation describing the scale dependence of the Green's functions:
\begin{equation}
    \left(\sum_{\phi, k} p_{k}^{\phi} \frac{\partial}{\partial p_{k}^{\phi}}+(m-\beta_m)\frac{\partial}{\partial m}-\beta_{g} \frac{\partial}{\partial g}+\sum_{\phi}n_{\phi}\left(c_{\phi}-\gamma_\phi \right)-4\right) G^{\left(n_{\phi}\right)}\left(p_{k}^{\phi}\right)=0.
    \label{equ:QCD-running}
\end{equation}
This equation implies that the couplings run with momentum or the characteristic energy of the process:
\begin{equation}
    \begin{aligned}
        \frac{d}{d \log (p / \mu)} \bar{g}(p ; g)&=\beta(\bar{g}), \quad \bar{g}(p=\mu ; g)=g, \\
        \frac{d}{d \log (p / \mu)} \bar{m}(p ; m)&=\beta(\bar{m}), \quad \bar{m}(p=\mu ; m)=m.
    \end{aligned}
\end{equation}
In the subsequent analysis, we will show how an analogous derivation in the scale-invariant theory with a dilaton leads to identical running behavior in the broken phase.

In the scale-invariant extension of QCD constructed in the previous section, several important modifications arise in the treatment of renormalized n-point Green's function. The field content now includes the dilaton, so the momentum space version of the Ward identity becomes 
\begin{equation}
    \left(\sum_{\phi, k} p_{k}^{\phi} \frac{\partial}{\partial p_{k}^{\phi}}+f_\chi\frac{\partial}{\partial f_\chi}+\sum_{\phi}\left(c_{\phi} n_{\phi}\right)-4\right) G^{\left(n_{\phi}\right)}\left(p_{k}^{\phi}\right)=0,
    \label{equ:QCDdilaton-ward}
\end{equation}
where $\phi=(G_\mu,\psi,\bar{\psi},\chi)$ with $c_\phi=(3,\frac{5}{2},\frac{5}{2},3)$, and we have explicitly included the derivative with respect to the VEV $f_\chi$. A second important change is that the renormalization scale is now dependent on the VEV through the dimensionless parameter $z$, defining  $\mu_0=zf_\chi$. Physical observables must be independent of the arbitrary choice of $z$, leading to a Callan-Symanzik-like equation with respect to $z$:
\begin{equation}
    \left(z \frac{\partial}{\partial z}+\hat{\beta}_{g} \frac{\partial}{\partial g}+\hat{\beta}_{y} \frac{\partial}{\partial y}+\sum_{\phi} \hat{\gamma}_{\phi} n_{\phi}-\hat{\gamma}_{\chi} f_\chi \frac{\partial}{\partial f_\chi}\right) G^{\left(n_\phi\right)}\left(p_{k}^{\phi}\right)=0,
    \label{equ:QCDdilaton-CS}
\end{equation}
where the hatted beta functions and anomalous dimensions are now defined as
\begin{equation}
    \hat{\beta}_g=z\frac{\partial g}{\partial z},\quad \hat{\beta}_y=z\frac{\partial y}{\partial z},\quad \hat{\gamma}_\phi=\frac{z}{2Z_\phi}\frac{\partial Z_\phi}{\partial z}.
    \label{defofbetafunc}
\end{equation}
Here, the dimensional fermion mass $m$ of standard QCD is replaced by the dimensionless Yukawa coupling $y$, and the $z$ dependence of the VEV $f_\chi$ is given by the field renormalization of the dilaton (consistent with the counter-terms being identical to those in the symmetric phase under appropriate renormalization schemes~\cite{Collins:1984xc}). Finally, in order to eliminate the explicit dependence on $f_\chi$, we use the quantum action principle, which yields the additional scaling relation
\begin{equation}
    \left(f_\chi\frac{\partial}{\partial f_\chi}-z\frac{\partial}{\partial z}-y\frac{\partial}{\partial y}\right)G^{\left(n_{\phi}\right)}\left(p_{k}^{\phi}\right)=-i\left\langle\chi\frac{\partial\mathcal{L}}{\partial\chi}\prod_{k}\phi(p_k^\phi)\right\rangle.
    \label{equ:quantum-action-principle}
\end{equation}
The right-hand side corresponds to diagrams with a single insertion of the dilaton. We now consider the decoupling limit $y\rightarrow0$ while keeping the physical fermion mass $m=y f_\chi$ finite. 
In this limit, the dilaton becomes non-propagating, the anomalous dimension $\gamma_\chi$ vanishes at the relevant order, and the insertion term on the right-hand side of Eq.\,\eqref{equ:quantum-action-principle} can be neglected. Combining Eqs.\,\eqref{equ:QCDdilaton-ward}-\eqref{equ:quantum-action-principle} then produces a scaling equation of identical functional form to Eq.\,\eqref{equ:QCD-running}:
\begin{equation}
    \left(\sum_{\phi, k} p_{k}^{\phi}\right. \frac{\partial}{\partial p_{k}^{\phi}}-\hat{\beta}_{g} \frac{\partial}{\partial g}+\left(y-\hat{\beta}_{y}\right) \frac{\partial}{\partial y}+\left.\sum_{\phi}n_{\phi}\left(c_{\phi}-\hat{\gamma}_{\phi} \right)-4\right) G^{\left(n_{\phi}\right)}\left(p_{k}^{\phi}\right)=0,
\label{equ:QCDdilaton-running}
\end{equation}
with the operator $m\frac{\partial}{\partial m}$ being replaced by $y\frac{\partial}{\partial y}$. 

The beta functions $\beta_g,\beta_m$ in standard QCD and their counterparts $\hat{\beta}_g,\hat{\beta}_y$ in the scale-invariant theory are both determined by the $1/\epsilon$ poles in dimensional regularization that renormalize their respective couplings. The effective Lagrangian of our scale-invariant formulation in Eq.\,\eqref{equ:LdilatonQCDd} differs from the standard QCD Lagrangian in Eq.\,\eqref{equ:LQCD} by an evanescent operator of $\mathcal{O}(\epsilon)$ and by dilaton-mediated interactions proportional to the small Yukawa coupling $y$. In the decoupling limit, these additional terms do not contribute to the one-loop divergences. Consequently, the one-loop beta functions calculated in the scale-invariant framework coincide exactly with those of standard QCD. Differences emerge only starting at the two-loop level, where evanescent interactions can contribute to the renormalization group flow through finite $1/\epsilon$ poles. However, this difference is controlled by dilaton interactions that are suppressed either by $1/f_\chi$ or by the small coupling $y = m_f/f_\chi$. Consequently, such higher-order effects on the dilaton couplings driven by the $\beta$-functions are suppressed by inverse powers of $f_\chi$ and thus negligible.

We conclude that the theory with spontaneously broken scale invariance reproduces the standard one-loop running of couplings in the broken phase, with the conventional renormalization scale dependence $\ln\mu$ being replaced by $\ln (z f_\chi)$. This equivalence not only demonstrates the quantum consistency of the formalism, but also highlights its utility as a powerful top-down approach for constructing dilaton EFTs where scale symmetry is broken spontaneously. 

\section{The Dilaton Linear Lagrangian}
\label{sec:DilatonEFTs}
Having established the framework for the dilaton effective theory, we now apply it to the SM to derive the low-energy dilaton-mediated interactions. The spontaneous breaking of scale symmetry occurs at the new physics (NP) scale $\Lambda$, determined by the dilaton VEV $f_\chi=\langle\Phi\rangle$. All new degrees of freedom responsible for this breaking are assumed to be significantly heavier than $\Lambda$ and can be safely integrated out.

Since $\Lambda$ is taken to lie well above the electroweak scale $v\simeq 246$ GeV, the theory just below $\Lambda$ consists of the SM fields plus the dilaton, described by an effective Lagrangian that respects both $SU(3)_c\times SU(2)_L\times U(1)_Y$ and approximate scale invariance. After scale symmetry breaking, the interactions among the SM fields themselves are encoded within the SMEFT framework, while the universal dilaton couplings to the SM trace anomaly provide the leading dilaton-mediated interactions.

To analyze dilaton phenomenology at colliders and in low-energy measurements, we derive the effective Lagrangian at different energy scales. This is achieved by renormalization group evolving the full $SU(3)_c\times SU(2)_L\times U(1)_Y$-invariant Lagrangian (including the dilaton) down to the electroweak scale and subsequent matching onto an $SU(3)_c\times U(1)_\mathrm{em}$-invariant effective field theory after electroweak symmetry breaking. In the following, we adopt the manifestly scale-invariant regularization framework, which allows a systematic treatment of renormalization group effects in the dilaton couplings.

\subsection{Scale-invariant SM}
\label{sec:scaleinvariantSM}
The SM Lagrangian is classically scale-invariant, with the exception of the negative quadratic term in the Higgs potential which explicitly breaks the symmetry. To restore scale invariance, the dilaton field $\Phi$ is introduced:  
\begin{equation}
\begin{aligned}
\mathcal{L}_\mathrm{inv}^{(4)}= & -\frac{1}{4g_3^2} G_{\mu \nu}^{A} G^{A \mu \nu}-\frac{1}{4g_2^2} W_{\mu \nu}^{I} W^{I \mu \nu}-\frac{1}{4g_1^2} B_{\mu \nu} B^{\mu \nu}+\sum_{\psi=Q, u, d, L, e} \bar{\psi} i \slashed{D}  \psi -\mathcal{L}_\mathrm{yuk} \\
& +\left(D_{\mu} H\right)^{\dagger}\left(D^{\mu} H\right)+\frac{1}{2}\partial_\mu\Phi\partial^\mu\Phi - V_0,\\
\mathcal{L}_\mathrm{yuk}=&\left(\bar{Q}H Y_{d}d+\bar{Q}\Tilde{H} Y_{u} u+\bar{L}H Y_{e}e+\text{h.c.}\right),
\end{aligned}
\label{SM4}
\end{equation}
where the most general, scale-invariant potential involving the Higgs doublet and the scalar field is given by~\cite{Ghilencea:2016dsl}
\begin{equation}
\begin{aligned}
V_0=\frac{\lambda_\phi}{3!}(H^\dagger H)^2+\frac{\lambda_m}{2}(H^\dagger H)\Phi^2+\frac{\lambda_\chi}{4!}\Phi^4,\quad H=\binom{G^{+}}{\frac{1}{\sqrt{2}}\left(\phi+i G^{0}\right)}.
\end{aligned}
\end{equation}
The terms proportional to $\lambda_\phi$ and $\lambda_\chi$ describe the self-interactions of the Higgs and the dilaton respectively, while the term proportional to $\lambda_m$ acts as the portal interaction responsible for generating the Higgs mass upon spontaneous breaking of scale symmetry. To obtain a charge-neutral VEV, we consider the potential involving the neutral components
\begin{equation}
    V_0(\phi,\Phi)=\frac{\lambda_\phi}{4!}\phi^4+\frac{\lambda_m}{4}\phi^2\Phi^2+\frac{\lambda_\chi}{4!}\Phi^4.
\end{equation}
Spontaneous symmetry breaking requires $V_0$ to develop a non-trivial vacuum with nonzero expectation values $v\equiv\langle\phi\rangle$ and $f_\chi\equiv\langle\Phi\rangle$, where $v\simeq246$~GeV is fixed by electroweak phenomenology. The minimization conditions
\begin{equation}
    \left.\partial_\phi V_0\right|_{v,f_\chi}=0,\quad\left.\partial_\Phi V_0\right|_{v,f_\chi}=0,
\end{equation}
lead to the relations 
\begin{equation}
    \lambda_\chi\lambda_\phi=9\lambda_m^2,\quad \frac{v^2}{f_\chi^2}=\frac{-3\lambda_m}{\lambda_\phi}=\frac{\lambda_\chi}{-3\lambda_m}.
\end{equation} 
These conditions guarantee the presence of a classically flat direction in the field space. Crucially, manifestly scale-invariant regularization schemes preserve this flatness at the quantum level, preventing the standard Coleman-Weinberg mechanism~\cite{Coleman:1973jx} from triggering scale symmetry breaking. Instead, spontaneous symmetry breaking may occur through alternative means such as inertial breaking~\cite{Ferreira:2018itt}, where the vacuum is selected along the flat direction by external factors such as cosmological initial conditions or non-minimal couplings to gravity. Finally, vacuum stability requires $\lambda_\phi>0$ and $\lambda_\chi>0$, while the phenomenological hierarchy $v\ll f_\chi$ implies the coupling relation $\lambda_\chi\ll-3\lambda_m\ll\lambda_\phi$.

\subsection{Dilaton couplings to the SM}
\label{sec:SMEFT}
Following the scale-invariant regularization framework from the previous section, we analytically continue to $d=4-2\epsilon$ dimensions and expand the dilaton field $\Phi$ around its VEV after spontaneous scale symmetry breaking. The resulting effective Lagrangian is given by
\begin{equation}
\begin{aligned}   
    \mathcal{L}_\text{eff}&=\mathcal{L}_\mathrm{SM}+\mathcal{L}_{\chi,\text{SM}},\\
    \mathcal{L}_\mathrm{SM}&=-\frac{1}{4}G_{\mu \nu}^{A} G^{A \mu \nu}-\frac{1}{4}W_{\mu \nu}^{I} W^{I \mu \nu}-\frac{1}{4}B_{\mu \nu} B^{\mu \nu}+\left(D_{\mu} H\right)^{\dagger}\left(D^{\mu} H\right)-\left(\frac{\lambda_\phi}{4!}\phi^4+\frac{\lambda_m}{4}f_\chi^2\phi^2\right)\\
    &+\sum_{\psi}\bar{\psi}i\slashed{D}\psi-\left(\bar{d}\frac{\phi}{\sqrt{2}} y_{d}d+\bar{u}\frac{\phi}{\sqrt{2}} y_{u} u+\bar{e}\frac{\phi}{\sqrt{2}} y_{e}e+\text{h.c.}\right),\\
    \mathcal{L}_{\chi,\text{SM}}&=-\frac{\chi}{f_\chi}\left\{\left(\frac{\beta_3}{2g_3}G^2+\frac{\beta_2}{2g_2}W^2+\frac{\beta_1}{2g_1}B^2\right)-\left(\frac{\beta_\phi}{4!}\phi^4+\left(\beta_m-2\lambda_m\right)\frac{f_\chi^2}{4}\phi^2\right)\right.\\
    &\left.-\left(\bar{d}\frac{\phi}{\sqrt{2}} \beta_{d}d+\bar{u}\frac{\phi}{\sqrt{2}} \beta_{u} u+\bar{e}\frac{\phi}{\sqrt{2}} \beta_{e}e+\text{h.c.}\right)\right\},
\end{aligned}
\label{equ:dilaton-SM-before-mixing}
\end{equation}
where $\mathcal{L}_\mathrm{SM}$ corresponds to the full SM Lagrangian, with the Higgs quadratic term generated proportionally to the VEV $f_\chi$ and $\mathcal{L}_{\chi,\text{SM}}$ describes the interaction between the dilaton $\chi$ and the SM fields. The dilaton couplings are determined by the trace anomaly and are thus proportional to the beta functions, which coincide with those of the conventional SM at one-loop level, as required by our formalism. 

Using the known renormalization-group evolution of the couplings, we evolve the effective Lagrangian down to the weak scale $\mu_w$, where electroweak symmetry is spontaneously broken. In the broken phase, it is appropriate to express the dilaton interactions in the mass-eigenstate basis of the physical fields.

For the gauge boson sector, this requires rotating to the physical $W^\pm,Z$ and photon fields $F$. For the fermions, it involves unitary transformations that diagonalize the Yukawa matrices, e.g., $U_u^{\dagger} Y_u R_{u} = Y_u^{\text{diag}}$. In the scalar sector, $h$ and $\chi$ mix to form mass eigenstates $\tilde{h}$ and $\tilde{\chi}$, 
\begin{equation}
    \begin{aligned}
        \tilde{h}&=\cos\alpha h+\sin\alpha\chi,\\
        \tilde{\chi}&=-\sin\alpha h+\cos\alpha\chi,
    \end{aligned}
\end{equation}
where the mixing angle $\alpha$ diagonalizes the scalar mass-squared matrix and satisfies $\tan^2\alpha=v^2/f_\chi^2$. For notational convenience, we henceforth drop the tildes and redefine $h$ and $\chi$ to denote their respective physical mass eigenstates. We identify $h$ as the SM Higgs boson observed at the LHC with a mass $m_h=125$~GeV, and $\chi$ as the massless pseudo-Goldstone dilaton. Therefore, in the physical mass-eigenstate basis, the dilaton $\chi$ couples to the SM fields both via the trace anomaly and through the scalar mixing with the Higgs. The complete dilaton-mediated interactions incorporating both effects are given by
\begin{equation}
    \begin{aligned}
        \mathcal{L}_\chi=&\mathcal{L}_\mathrm{\chi,g}+\mathcal{L}_\mathrm{\chi,\psi},\\
        \mathcal{L}_\mathrm{\chi,g}=&-\frac{\beta_{3}}{2 g_3}\frac{\chi}{f_\chi} G_{\mu \nu}^{A} G^{A \mu \nu}-\frac{\beta_e}{2e} \frac{\chi}{f_\chi}2c_{WW} W^+_{\mu\nu}W^{-\mu\nu}+2\frac{\chi}{f_\chi}m_W^2W_\mu^-W^{+\mu}\\
    &-\frac{\beta_e}{2e} \frac{\chi}{f_\chi}\left(c_{ZZ} Z_{\mu\nu}Z^{\mu\nu}+2c_{\gamma Z} F_{\mu\nu}Z^{\mu\nu}
    +c_{\gamma\gamma} F_{\mu\nu}F^{\mu\nu}\right)+\frac{\chi}{f_\chi}m_Z^2Z_\mu Z^\mu,\\
    \mathcal{L}_\mathrm{\chi,\psi}=&-\sum_{\psi=u,d,e}\left(1+\gamma_{m_\psi}\right)m_\psi\frac{\chi}{f_\chi}\bar{\psi}\psi,
    \end{aligned}
\end{equation}
where $m_W=g_2v/2$, $m_Z=\sqrt{g_2^2+g_1^2}v/2$ and $m_\psi=Y_\psi^\mathrm{diag}v/\sqrt{2}$ are the conventional SM masses. Crucially, the dilaton-Higgs mixing is responsible for the terms explicitly proportional to the standard mass scales: the $m_W^2$ and $m_Z^2$ terms for the vector bosons, and the leading unit contribution ($1\cdot m_\psi$) to the fermion couplings. The remaining terms, specifically the fermion contribution proportional to the anomalous dimension $\gamma_{m_\psi}$ and the field strength tensor interactions, originate from the trace anomaly. The coefficients $c_{ij}$ are determined by the beta functions and the weak mixing angle $\theta_w$
\begin{equation}
\label{equ:cij_formula}
    c_{WW}=\frac{\beta_2}{\beta_e}s_w,\quad c_{ZZ}=\frac{\beta_2}{\beta_e}c_w^2s_w+\frac{\beta_1}{\beta_e}s_w^2c_w,\quad c_{\gamma Z}=\frac{\beta_2}{\beta_e}s_w^2c_w-\frac{\beta_1}{\beta_e}c_w^2s_w,\quad c_{\gamma\gamma}=\frac{\beta_2}{\beta_e}s_w^3+\frac{\beta_1}{\beta_e}c_w^3,
\end{equation}
with $s_w=\sin\theta_w$ and $c_w=\cos\theta_w$. The effective Lagrangian $\mathcal{L}_\chi$ at the weak scale, with all parameters defined at $\mu\gtrsim \mu_w$, provides the foundation for our subsequent phenomenological analysis.

The above discussion assumes exact scale invariance. Through the scale-invariant regularization scheme, the dilaton remains exactly massless even at the quantum level after spontaneous scale symmetry breaking~\cite{Ghilencea:2016dsl,Oda:2021rcj,Nogradi:2021zqw}. When explicit scale-symmetry breaking effects are present, the dilaton acquires a mass accordingly. A general analysis can be carried out by introducing a nearly marginal operator of scaling dimension $4-\epsilon$ to the initial scale-invariant theory~\cite{PhysRevD.87.115006}
\begin{equation}
    \mathcal{L}=\mathcal{L}_\text{SI}+\lambda_\mathcal{O}\mathcal{O}(x).
\end{equation}
The renormalization group evolution of the coupling $\lambda_\mathcal{O}$ determines the scaling of this operator. After spontaneous scale symmetry breaking, the resulting dilaton potential leads to a mass $m_\chi\propto\sqrt{\epsilon} f_\chi$. Corrections to the effective Lagrangian from a non-zero dilaton mass are of order $\mathcal{O}(m_\chi^2/f_\chi^2)$ and are neglected in the following analysis.

\subsection{Higher-dimensional operators}
\label{sec:Higher-dimensional operators}
It is straightforward to extend the scale-invariant framework beyond the renormalizable terms of the SM to include higher-dimensional operators. A natural illustration can be provided by the dimension-$5$ Weinberg operator $\frac{c^{(5)}}{\Lambda}L^T\tilde{H}^*C^\dagger\tilde{H}^\dagger L$~\cite{Weinberg:1979sa, Abdullahi:2022jlv}, which generates Majorana neutrino masses.

To construct its scale-invariant counterpart, we consider the Type-I seesaw mechanism as a representative UV completion. 
For simplicity, we consider a single generation of leptons and the right-handed neutrino. In $d=4-2\epsilon$ dimensions, the relevant scale-invariant mass terms are
\begin{equation}
\mathcal{L}_\text{m}=-\mu^\epsilon(\Phi)Y_{\alpha } \bar{L}_{\alpha} \tilde{H}N -\mu^\epsilon(\Phi)y\Phi\overline{{N}^{c}} N+\text { h.c. },
\end{equation}
where $Y_\alpha$ are the Dirac-neutrino Yukawa couplings of the SM Higgs doublet $H$ to the lepton doublet $L_\alpha$, and $y$ is the Majorana-neutrino Yukawa coupling of the dilaton field $\Phi$ to the right-handed neutrino $N$. After spontaneous scale symmetry breaking, $\langle\Phi\rangle=f_\chi$, generating a heavy Majorana mass $M_N=yf_\chi$. Integrating out $N$ at energies $\ll M_N$, substituting the classical equation of motion $N\approx -(Y/y\Phi)C(\bar{L}\tilde{H})^T$ back into the Lagrangian directly produces
\begin{equation}
\begin{aligned}
    \mathcal{L}^{(5)}_\text{eff}&=\frac{Y^2}{y\Phi}(\bar{L}\tilde{H})C(\bar{L}\tilde{H})^T+\mathrm{h.c.}\\
    &=\frac{Y^2}{yf_\chi}(\bar{L}\tilde{H})C(\bar{L}\tilde{H})^T-\frac{Y^2}{yf_\chi}\frac{\chi}{f_\chi}(\bar{L}\tilde{H})C(\bar{L}\tilde{H})^T+\mathrm{h.c.},
\end{aligned}
\end{equation}
where the first term matches the standard Weinberg operator, identifying $\Lambda\sim f_\chi$, and the second describes dilaton-neutrino interactions.

This result from the seesaw model is not a coincidence but reveals a powerful and general principle for constructing scale-invariant effective theories. In the absence of fundamental mass scales, all dimensional parameters arise dynamically from the dilaton VEV after spontaneous scale symmetry breaking. 
Therefore, any higher-dimensional operator in the SMEFT, conventionally written as $\mathcal{O}^{(d)}/\Lambda^{d-4}$, acquires a scale-invariant form
\begin{equation}
    \mathcal{L}_{\text{SMEFT,SI}}=\sum_ic_i\frac{\mathcal{O}_i^{d_i}}{\Phi^{d_i-4}},
    \label{eq:principle}
\end{equation}
where $c_i$ are the respective coupling constants and every power of the suppression scale $\Lambda$ in the denominator has been replaced by the same power of the dilaton field $\Phi$. 

\subsection{Low-Energy Dilaton Effective theory}
\label{sec:LEFT}
For a dilaton significantly lighter than the weak scale, it can mediate novel signatures in low-energy processes at typical energies $E\ll100$GeV. In this regime, the heavy degrees of freedom, the top quark, the Higgs boson and the weak gauge bosons $W^\pm$ and $Z$, are integrated out to obtain the corresponding low-energy effective theory. At tree-level matching, the Lagrangian for the dilaton is given by
\begin{equation}
    \mathcal{L}_{\text{QCD+QED},\chi}=-\frac{\chi}{f_\chi}\left[\frac{\beta_3}{2g_3}G^A_{\mu\nu}G^{A,\mu\nu}+\frac{\beta_e}{2e}c_{\gamma\gamma}F_{\mu\nu}F^{\mu\nu}+\sum_{\psi=u,d,e}\left(1+\gamma_{m_\psi}\right)m_\psi\bar{\psi}\psi\right],
    \label{equ:QCD+QED+dilaton}
\end{equation}
where $\beta_3$ and $\beta_e$ are the one-loop QCD and QED beta functions and $\gamma_{m_\psi}$ are the fermion mass anomalous dimensions. All coefficients, including $c_{\gamma\gamma}$, are evaluated at the matching scale.

Alternatively, these interactions can be parameterized by the LEFT~\cite{Jenkins:2017jig,Jenkins:2017dyc,Liao:2020zyx,Li:2020tsi,Murphy:2020cly} model-independently. In this section, we discuss the LEFT with dilaton and construct the interactions up to dimension 7.

\begin{table}[htbp]
\renewcommand\arraystretch{1.5}
    \centering
    \begin{tabular}{c|c|c|c|c|c}
\hline
\multirow{2}{*}{building block} & \multicolumn{2}{c|}{gauge symmetry} & \multicolumn{2}{c|}{global symmetry} & \multirow{2}{*}{scale symmetry} \\
\cline{2-5}
& $SU(3)_c$ & $U(1)_e$ & $U(1)_B$ & $U(1)_L$ & \\
\hline
$G^A_{\mu\nu}$ & $\mathbf{8}$ & $0$ & $0$ & $0$ & $e^{2\lambda}$ \\
\hline
$F_{\mu\nu}$ & $\mathbf{1}$ & $0$ & $0$ & $0$ & $e^{2\lambda}$ \\
\hline
$e$ & $\mathbf{1}$ & $-1$ & $0$ & $1$ & $e^{\frac{3}{2}\lambda}$ \\
\hline
$\nu$ & $\mathbf{1}$ & $0$ & $0$ & $1$ & $e^{\frac{3}{2}\lambda}$ \\
\hline
$u$ & $\mathbf{3}$ & $\frac{2}{3}$ & $\frac{1}{3}$ & $0$ & $e^{\frac{3}{2}\lambda}$ \\
\hline
$c$ & $\mathbf{3}$ & $\frac{2}{3}$ & $\frac{1}{3}$ & $0$ & $e^{\frac{3}{2}\lambda}$ \\
\hline
$d$ & $\mathbf{3}$ & $-\frac{1}{3}$ & $\frac{1}{3}$ & $0$ & $e^{\frac{3}{2}\lambda}$ \\
\hline
$s$ & $\mathbf{3}$ & $-\frac{1}{3}$ & $\frac{1}{3}$ & $0$ & $e^{\frac{3}{2}\lambda}$ \\
\hline
$b$ & $\mathbf{3}$ & $-\frac{1}{3}$ & $\frac{1}{3}$ & $0$ & $e^{\frac{3}{2}\lambda}$ \\
\hline
$\Phi$ & $\mathbf{1}$ & $0$ & $0$ & $0$ & $e^{\lambda}$ \\
\hline
    \end{tabular}
    \caption{The building blocks of the LEFT with the extension of the dilaton $\Phi$. The representations of the fields under the gauge symmetry and the scale symmetry are listed. Besides, two global symmetries $U(1)_B$ and $U(1)_L$ related to the baryon number and the lepton number are presented, which are nearly conserved in the LEFT. 
    To keep the gauge invariance, we use the field-strength tensors of the two gauge bosons as the building blocks.}
    \label{tab:building_blocks_LEFT}
\end{table}

The gauge group of this extended LEFT is the one of the QCD and the QED theories,
\begin{equation}
    \mathcal{G}_{\text{LEFT}} = \mathcal{G}_{\text{QCD}} \times \mathcal{G}_{\text{QED}} = SU(3)_c \times U(1)_{e}\,,
\end{equation}
which guarantees color conservation and electric charge conservation. The building blocks are the two associated gauge bosons, the gluon $G$ and the photon $A$, the leptons, and the light quarks. We present the building blocks and their representations in Tab.~\ref{tab:building_blocks_LEFT}, where the baryon number symmetry group $U(1)_B$ and the lepton number symmetry group $U(1)_L$ are also given. In particular, we use the field strength tensors of the gauge bosons as the building blocks to maintain the gauge invariance,
\begin{align}
    G^A_{\mu\nu} &= \partial_\mu G^A_\nu - \partial_\nu G^A_\mu + i f^{ABC} G^B_\mu G^B_\nu \,, \\
    F_{\mu\nu} &= \partial_\mu A_\nu - \partial_\nu A_\mu\,.
\end{align}

Similar to Eq.\,\eqref{eq:principle}, the effective Lagrangian of the LEFT extended by the dilaton should take the form 
\begin{align}
    \mathcal{L}_{\text{LEFT}} = \sum_{d=3} c^{(d)}_a \frac{\mathcal{O}^{(d)}_a}{\Phi^{d-4}}\,,
\end{align}
where $c_a^{(d)}$ are the dimensionless Wilson coefficients, and $\mathcal{O}^{(d)}_a$ are the independent operators of $d$-dimension without the dilaton. According to the expansion form of the dilaton
\begin{equation}
    \Phi = f_\chi+\chi\,,
\end{equation}
the Lagrangian can be divided into two sectors,
\begin{equation}
    \mathcal{L}_{\text{LEFT}} = \mathcal{L}_{\text{LEFT},0} + \mathcal{L}_{\text{LEFT},\chi\,},
\end{equation}
where $\mathcal{L}_{\text{LEFT},0}$ is the LEFT Lagrangian without the dilaton,
\begin{equation}
    \mathcal{L}_{\text{LEFT},0}= \sum_{d=3}\frac{c_a^{(d)}}{f_\chi^{d-4}}\mathcal{O}_a^{(d)}\,, 
\end{equation}
while the Lagrangian $\mathcal{L}_{\text{LEFT},\chi\,}$ contains the operators with the dilaton, which can be expressed as an expansion in terms of the fluctuation field $\chi$. Keeping the operators linear in the $\chi$, the Lagrangian $\mathcal{L}_{\text{LEFT},\chi\,}$ takes the form 
\begin{equation}
    \mathcal{L}_{\text{LEFT},\chi\,} = \sum_{d=3}\frac{(4-d)c_a^{(d)}}{f_\chi^{d-4}}\frac{\chi}{f_\chi}\mathcal{O}_a^{(d)}\,,
\end{equation}
which implies for every specific dimension $d$, the fluctuation field $\chi$ interacts with the other fields in the LEFT in a universal way.

Accordingly, we can list the effective operators in $\mathcal{L}_\chi$ linear in $\chi$. Next, we present these operators up to dimension 7.

\paragraph{Dimension-4, 5, 6}

\begin{center}
    \renewcommand{\arraystretch}{1.51}
    \begin{minipage}[t]{0.3\textwidth}
    \vspace{0pt}
\begin{tabular}{c|c}
        \hline
\multicolumn{2}{c}{$\boldsymbol{\chi(\nu\nu)}$} \\
\hline
$\O_{\chi\nu}$ & $\chi(\nu_{Lp}^T C \nu_{Lr})$  \\
\hline
\multicolumn{2}{c}{$\boldsymbol{\chi(\overline{L}R)}$} \\
\hline
$\O_{\chi e}$ & $\chi \bar e_{Lp}    e_{Rr}$  \\
$\O_{\chi u }$ & $\chi \bar u_{Lp}     u_{Rr}$   \\
$\O_{\chi d}$ & $\chi \bar d_{Lp}   d_{Rr}$  \\
        \hline
    \end{tabular}
    \end{minipage}
    \begin{minipage}[t]{0.3\textwidth}
    \vspace{0pt}
        \begin{tabular}{c|c}
\hline
\multicolumn{2}{c}{$\boldsymbol{\chi(\nu\nu)X+\hc}$} \\
\hline
$\O_{\chi \nu \gamma}$ & $ \chi (\nu_{Lp}^T C   \sigma^{\mu \nu}  \nu_{Lr})  F_{\mu \nu} $ \\
\hline
\multicolumn{2}{c}{$\boldsymbol{\chi(\overline{L} R)X+\hc}$} \\
\hline
$\O_{\chi e \gamma}$ & $ \chi \bar e_{Lp}   \sigma^{\mu \nu} e_{Rr}\, F_{\mu \nu}$  \\
$\O_{\chi u \gamma}$ & $ \chi \bar u_{Lp}   \sigma^{\mu \nu}  u_{Rr}\, F_{\mu \nu}$   \\
$\O_{\chi d \gamma}$ & $ \chi \bar d_{Lp}  \sigma^{\mu \nu} d_{Rr}\, F_{\mu \nu}$  \\
$\O_{\chi u G}$ & $ \chi \bar u_{Lp}   \sigma^{\mu \nu}  T^A u_{Rr}\,  G_{\mu \nu}^A$  \\
$\O_{\chi d G}$ & $ \chi \bar d_{Lp}   \sigma^{\mu \nu} T^A d_{Rr}\,  G_{\mu \nu}^A$ \\
\hline
        \end{tabular}
    \end{minipage}
    
    \begin{minipage}[t]{0.3\textwidth}
    \vspace{0pt}
        \begin{tabular}{c|c}
        \hline
\multicolumn{2}{c}{$\boldsymbol{\chi X^2}$} \\
\hline
$\O_{\chi G}$ & $\chi G^A_{\mu\nu}G^{A}{}^{\mu\nu} $ \\
        \hline
    \end{tabular}
    \end{minipage}
\end{center}

\paragraph{Dimension-7}
    \begin{center}
        \renewcommand{\arraystretch}{1.51}
\begin{minipage}[t][10cm][t]{7cm}
    \vspace{0pt}
    \begin{tabular}{c|c}
\hline
\multicolumn{2}{c}{$\boldsymbol{\chi (\overline{L}L)(\overline{L}L)}$} \\
\hline
$\op{\chi\nu\nu}{V}{LL}$ & $\chi(\bar \nu_{Lp}  \gamma^\mu \nu_{Lr} )(\bar \nu_{Ls} \gamma_\mu \nu_{Lt})$   \\
$\op{\chi ee}{V}{LL}$       & $\chi(\bar e_{Lp}  \gamma^\mu e_{Lr})(\bar e_{Ls} \gamma_\mu e_{Lt})$   \\
$\op{\chi \nu e}{V}{LL}$       & $\chi(\bar \nu_{Lp} \gamma^\mu \nu_{Lr})(\bar e_{Ls}  \gamma_\mu e_{Lt})$  \\
$\op{\chi\nu u}{V}{LL}$       & $\chi(\bar \nu_{Lp} \gamma^\mu \nu_{Lr}) (\bar u_{Ls}  \gamma_\mu u_{Lt})$  \\
$\op{\chi\nu d}{V}{LL} $      & $\chi(\bar \nu_{Lp} \gamma^\mu \nu_{Lr})(\bar d_{Ls} \gamma_\mu d_{Lt})$     \\
$\op{\chi eu}{V}{LL}$      & $\chi(\bar e_{Lp}  \gamma^\mu e_{Lr})(\bar u_{Ls} \gamma_\mu u_{Lt})$   \\
$\op{\chi ed}{V}{LL}$       & $\chi(\bar e_{Lp}  \gamma^\mu e_{Lr})(\bar d_{Ls} \gamma_\mu d_{Lt})$  \\
$\op{\chi\nu edu}{V}{LL}$      &$ \chi(\bar \nu_{Lp} \gamma^\mu e_{Lr}) (\bar d_{Ls} \gamma_\mu u_{Lt})  + \hc$   \\
$\op{\chi uu}{V}{LL}$        & $\chi(\bar u_{Lp} \gamma^\mu u_{Lr})(\bar u_{Ls} \gamma_\mu u_{Lt})$    \\
$\op{\chi dd}{V}{LL}$   & $\chi(\bar d_{Lp} \gamma^\mu d_{Lr})(\bar d_{Ls} \gamma_\mu d_{Lt})$    \\
$\op{\chi ud}{V1}{LL}$     & $\chi(\bar u_{Lp} \gamma^\mu u_{Lr}) (\bar d_{Ls} \gamma_\mu d_{Lt})$  \\
$\op{\chi ud}{V8}{LL}$     & $\chi(\bar u_{Lp} \gamma^\mu T^A u_{Lr}) (\bar d_{Ls} \gamma_\mu T^A d_{Lt})$   \\
\hline
    \end{tabular}
\end{minipage}
\begin{minipage}[t][1cm][t]{7cm}
    \vspace{0pt}
    \begin{tabular}{c|c}
\hline
\multicolumn{2}{c}{$\boldsymbol{\chi (\overline{R}R)(\overline{R}R)}$} \\
\hline
$\op{\chi ee}{V}{RR}$     & $\chi (\bar e_{Rp} \gamma^\mu e_{Rr})(\bar e_{Rs} \gamma_\mu e_{Rt})$  \\
$\op{\chi eu}{V}{RR}$       & $\chi (\bar e_{Rp}  \gamma^\mu e_{Rr})(\bar u_{Rs} \gamma_\mu u_{Rt})$   \\
$\op{\chi ed}{V}{RR}$     & $\chi (\bar e_{Rp} \gamma^\mu e_{Rr})  (\bar d_{Rs} \gamma_\mu d_{Rt})$   \\
$\op{\chi uu}{V}{RR}$      & $\chi (\bar u_{Rp} \gamma^\mu u_{Rr})(\bar u_{Rs} \gamma_\mu u_{Rt})$  \\
$\op{\chi dd}{V}{RR}$      & $\chi (\bar d_{Rp} \gamma^\mu d_{Rr})(\bar d_{Rs} \gamma_\mu d_{Rt})$    \\
$\op{\chi ud}{V1}{RR}$       & $\chi (\bar u_{Rp} \gamma^\mu u_{Rr}) (\bar d_{Rs} \gamma_\mu d_{Rt})$  \\
$\op{\chi ud}{V8}{RR}$    & $\chi (\bar u_{Rp} \gamma^\mu T^A u_{Rr}) (\bar d_{Rs} \gamma_\mu T^A d_{Rt})$  \\
\hline
\multicolumn{2}{c}{$\boldsymbol{\chi X^3}$} \\
\hline
$\O_{\chi G}$     & $\chi f^{ABC} G_\mu^{A\nu} G_\nu^{B\rho} G_\rho^{C\mu}$  \\
$\O_{\chi\widetilde G}$ & $\chi f^{ABC} \widetilde G_\mu^{A\nu} G_\nu^{B\rho} G_\rho^{C\mu}$  \\
\hline
    \end{tabular}
\end{minipage}
    \end{center}

\newpage
    \begin{center}
        \renewcommand{\arraystretch}{1.51}
        \begin{minipage}[t][1cm][t]{7cm}
            \vspace{0pt}
        \begin{tabular}{c|c}
\hline
\multicolumn{2}{c}{$\boldsymbol{\chi (\overline{L}L)(\overline{R}R)}$} \\
\hline
$\op{\chi \nu e}{V}{LR}$     & $\chi (\bar \nu_{Lp} \gamma^\mu \nu_{Lr})(\bar e_{Rs}  \gamma_\mu e_{Rt})$  \\
$\op{\chi ee}{V}{LR}$       & $\chi (\bar e_{Lp}  \gamma^\mu e_{Lr})(\bar e_{Rs} \gamma_\mu e_{Rt})$ \\
$\op{\chi \nu u}{V}{LR}$         & $\chi (\bar \nu_{Lp} \gamma^\mu \nu_{Lr})(\bar u_{Rs}  \gamma_\mu u_{Rt})$    \\
$\op{\chi \nu d}{V}{LR}$         & $\chi (\bar \nu_{Lp} \gamma^\mu \nu_{Lr})(\bar d_{Rs} \gamma_\mu d_{Rt})$   \\
$\op{\chi eu}{V}{LR}$        & $\chi (\bar e_{Lp}  \gamma^\mu e_{Lr})(\bar u_{Rs} \gamma_\mu u_{Rt})$   \\
$\op{\chi ed}{V}{LR}$        & $\chi (\bar e_{Lp}  \gamma^\mu e_{Lr})(\bar d_{Rs} \gamma_\mu d_{Rt})$   \\
$\op{\chi ue}{V}{LR}$        & $\chi (\bar u_{Lp} \gamma^\mu u_{Lr})(\bar e_{Rs}  \gamma_\mu e_{Rt})$   \\
$\op{\chi de}{V}{LR}$         & $\chi (\bar d_{Lp} \gamma^\mu d_{Lr}) (\bar e_{Rs} \gamma_\mu e_{Rt})$   \\
$\op{\chi \nu edu}{V}{LR}$        & $\chi (\bar \nu_{Lp} \gamma^\mu e_{Lr})(\bar d_{Rs} \gamma_\mu u_{Rt})  +\hc$ \\
$\op{\chi uu}{V1}{LR}$        & $\chi (\bar u_{Lp} \gamma^\mu u_{Lr})(\bar u_{Rs} \gamma_\mu u_{Rt})$   \\
$\op{\chi uu}{V8}{LR}$       & $\chi (\bar u_{Lp} \gamma^\mu T^A u_{Lr})(\bar u_{Rs} \gamma_\mu T^A u_{Rt})$    \\ 
$\op{\chi ud}{V1}{LR}$       & $\chi (\bar u_{Lp} \gamma^\mu u_{Lr}) (\bar d_{Rs} \gamma_\mu d_{Rt})$  \\
$\op{\chi ud}{V8}{LR}$       & $\chi (\bar u_{Lp} \gamma^\mu T^A u_{Lr})  (\bar d_{Rs} \gamma_\mu T^A d_{Rt})$  \\
$\op{\chi du}{V1}{LR}$       & $\chi (\bar d_{Lp} \gamma^\mu d_{Lr})(\bar u_{Rs} \gamma_\mu u_{Rt})$   \\
$\op{\chi du}{V8}{LR}$       & $\chi (\bar d_{Lp} \gamma^\mu T^A d_{Lr})(\bar u_{Rs} \gamma_\mu T^A u_{Rt})$ \\
$\op{\chi dd}{V1}{LR} $     & $\chi (\bar d_{Lp} \gamma^\mu d_{Lr})(\bar d_{Rs} \gamma_\mu d_{Rt})$  \\
$\op{\chi dd}{V8}{LR}$   & $\chi (\bar d_{Lp} \gamma^\mu T^A d_{Lr})(\bar d_{Rs} \gamma_\mu T^A d_{Rt})$ \\
$\op{\chi uddu}{V1}{LR}$   & $\chi (\bar u_{Lp} \gamma^\mu d_{Lr})(\bar d_{Rs} \gamma_\mu u_{Rt})  + \hc$  \\
$\op{\chi uddu}{V8}{LR}$      & $\chi (\bar u_{Lp} \gamma^\mu T^A d_{Lr})(\bar d_{Rs} \gamma_\mu T^A  u_{Rt})  + \hc$ \\
\hline
        \end{tabular}
    \end{minipage}
    \begin{minipage}[t][1cm][t]{7cm}
        \centering
        \vspace{0pt}
        \begin{tabular}{c|c}
            \hline
\multicolumn{2}{c}{$\boldsymbol{\chi (\overline{L}R)(\overline{L}R) + \hc}$} \\
            \hline
$\op{\chi ee}{S}{RR}$ 		& $\chi (\bar e_{Lp}   e_{Rr}) (\bar e_{Ls} e_{Rt})$   \\
$\op{\chi eu}{S}{RR}$  & $\chi (\bar e_{Lp}   e_{Rr}) (\bar u_{Ls} u_{Rt})$   \\
$\op{\chi eu}{T}{RR}$ & $\chi (\bar e_{Lp}   \sigma^{\mu \nu}   e_{Rr}) (\bar u_{Ls}  \sigma_{\mu \nu}  u_{Rt})$  \\
$\op{\chi ed}{S}{RR}$  & $\chi (\bar e_{Lp} e_{Rr})(\bar d_{Ls} d_{Rt})$  \\
$\op{\chi ed}{T}{RR}$ & $\chi (\bar e_{Lp} \sigma^{\mu \nu} e_{Rr}) (\bar d_{Ls} \sigma_{\mu \nu} d_{Rt})$   \\
$\op{\chi \nu edu}{S}{RR}$ & $\chi (\bar   \nu_{Lp} e_{Rr})  (\bar d_{Ls} u_{Rt} )$ \\
$\op{\chi \nu edu}{T}{RR}$ &  $\chi (\bar  \nu_{Lp}  \sigma^{\mu \nu} e_{Rr} )  (\bar  d_{Ls}  \sigma_{\mu \nu} u_{Rt} )$   \\
$\op{\chi uu}{S1}{RR}$  & $\chi (\bar u_{Lp}   u_{Rr}) (\bar u_{Ls} u_{Rt})$  \\
$\op{\chi uu}{S8}{RR}$   & $\chi (\bar u_{Lp}   T^A u_{Rr}) (\bar u_{Ls} T^A u_{Rt})$  \\
$\op{\chi ud}{S1}{RR}$   & $\chi (\bar u_{Lp} u_{Rr})  (\bar d_{Ls} d_{Rt})$   \\
$\op{\chi ud}{S8}{RR}$  & $\chi (\bar u_{Lp} T^A u_{Rr})  (\bar d_{Ls} T^A d_{Rt})$  \\
$\op{\chi dd}{S1}{RR}$   & $\chi (\bar d_{Lp} d_{Rr}) (\bar d_{Ls} d_{Rt})$ \\
$\op{\chi dd}{S8}{RR}$  & $\chi (\bar d_{Lp} T^A d_{Rr}) (\bar d_{Ls} T^A d_{Rt})$  \\
$\op{\chi uddu}{S1}{RR}$ &  $\chi (\bar u_{Lp} d_{Rr}) (\bar d_{Ls}  u_{Rt})$   \\
$\op{\chi uddu}{S8}{RR}$  &  $\chi (\bar u_{Lp} T^A d_{Rr}) (\bar d_{Ls}  T^A u_{Rt})$  \\
            \hline
        \end{tabular} \\
        \vspace{5pt}
        \begin{tabular}{c|c}
\hline
\multicolumn{2}{c}{$\boldsymbol{\chi (\overline{L}R)(\overline{R}L) + \hc}$} \\
\hline
$\op{\chi eu}{S}{RL}$  & $\chi (\bar e_{Lp} e_{Rr}) (\bar u_{Rs}  u_{Lt})$  \\
$\op{\chi ed}{S}{RL}$ & $\chi (\bar e_{Lp} e_{Rr}) (\bar d_{Rs} d_{Lt})$ \\
$\op{\chi \nu edu}{S}{RL}$  & $\chi (\bar \nu_{Lp} e_{Rr}) (\bar d_{Rs}  u_{Lt})$  \\
\hline
        \end{tabular}
    \end{minipage}
    \end{center}

\newpage
\begin{center}
    \renewcommand{\arraystretch}{1.51}
    \begin{minipage}[t]{7cm}
        \vspace{0pt}
        \begin{tabular}{c|c}
            \hline
\multicolumn{2}{c}{$\boldsymbol{\Delta L = 4 + \hc}$} \\
\hline
$\op{\chi \nu\nu}{S}{LL}$ &  $\chi (\nu_{Lp}^T C \nu_{Lr}^{}) (\nu_{Ls}^T C \nu_{Lt}^{} )$  \\
            \hline
        \end{tabular}\\\vspace{5pt}
        \begin{tabular}{c|c}
            \hline
\multicolumn{2}{c}{$\boldsymbol{\Delta L = 2 + \hc}$} \\
\hline
$\op{\chi \nu e}{S}{LL}$  &  $\chi (\nu_{Lp}^T C \nu_{Lr}) (\bar e_{Rs} e_{Lt})$   \\
$\op{\chi \nu e}{T}{LL}$ &  $\chi (\nu_{Lp}^T C \sigma^{\mu \nu} \nu_{Lr}) (\bar e_{Rs}\sigma_{\mu \nu} e_{Lt} )$  \\
$\op{\chi \nu e}{S}{LR}$ &  $\chi (\nu_{Lp}^T C \nu_{Lr}) (\bar e_{Ls} e_{Rt} )$  \\
$\op{\chi \nu u}{S}{LL}$  &  $\chi (\nu_{Lp}^T C \nu_{Lr}) (\bar u_{Rs} u_{Lt} )$  \\
$\op{\chi \nu u}{T}{LL}$  &  $\chi (\nu_{Lp}^T C \sigma^{\mu \nu} \nu_{Lr}) (\bar u_{Rs} \sigma_{\mu \nu} u_{Lt} )$ \\
$\op{\chi \nu u}{S}{LR}$  &  $\chi (\nu_{Lp}^T C \nu_{Lr}) (\bar u_{Ls} u_{Rt} )$  \\
$\op{\chi \nu d}{S}{LL}$   &  $\chi (\nu_{Lp}^T C \nu_{Lr}) (\bar d_{Rs} d_{Lt} )$ \\
$\op{\chi \nu d}{T}{LL}$   &  $\chi (\nu_{Lp}^T C \sigma^{\mu \nu}  \nu_{Lr}) (\bar d_{Rs} \sigma_{\mu \nu} d_{Lt} )$ \\
$\op{\chi \nu d}{S}{LR}$  &  $\chi (\nu_{Lp}^T C \nu_{Lr}) (\bar d_{Ls} d_{Rt} )$ \\
$\op{\chi \nu edu}{S}{LL}$ &  $\chi (\nu_{Lp}^T C e_{Lr}) (\bar d_{Rs} u_{Lt} )$  \\
$\op{\chi \nu edu}{T}{LL}$  & $\chi (\nu_{Lp}^T C  \sigma^{\mu \nu} e_{Lr}) (\bar d_{Rs}  \sigma_{\mu \nu} u_{Lt} )$ \\
$\op{\chi \chi \nu edu}{S}{LR}$   & $\chi (\nu_{Lp}^T C e_{Lr}) (\bar d_{Ls} u_{Rt} )$ \\
$\op{\chi \nu edu}{V}{RL}$   & $\chi (\nu_{Lp}^T C \gamma^\mu e_{Rr}) (\bar d_{Ls} \gamma_\mu u_{Lt} )$  \\
$\op{\chi \nu edu}{V}{RR}$   & $\chi (\nu_{Lp}^T C \gamma^\mu e_{Rr}) (\bar d_{Rs} \gamma_\mu u_{Rt} )$  \\
            \hline
        \end{tabular}
    \end{minipage}
    \begin{minipage}[t]{7cm}
        \vspace{0pt}
        \begin{tabular}{c|c}
            \hline
\multicolumn{2}{c}{$\boldsymbol{\Delta B=\Delta L = 1+ \hc}$} \\
\hline
$\op{\chi udd}{S}{LL}$ &  $\chi \epsilon_{\alpha\beta\gamma}  (u_{Lp}^{\alpha T} C d_{Lr}^{\beta}) (d_{Ls}^{\gamma T} C \nu_{Lt}^{})$   \\
$\op{\chi duu}{S}{LL}$ & $\chi \epsilon_{\alpha\beta\gamma}  (d_{Lp}^{\alpha T} C u_{Lr}^{\beta}) (u_{Ls}^{\gamma T} C e_{Lt}^{})$  \\
$\op{\chi uud}{S}{LR}$ & $\chi \epsilon_{\alpha\beta\gamma}  (u_{Lp}^{\alpha T} C u_{Lr}^{\beta}) (d_{Rs}^{\gamma T} C e_{Rt}^{})$  \\
$\op{\chi duu}{S}{LR}$ & $\chi \epsilon_{\alpha\beta\gamma}  (d_{Lp}^{\alpha T} C u_{Lr}^{\beta}) (u_{Rs}^{\gamma T} C e_{Rt}^{})$   \\
$\op{\chi uud}{S}{RL}$ & $\chi \epsilon_{\alpha\beta\gamma}  (u_{Rp}^{\alpha T} C u_{Rr}^{\beta}) (d_{Ls}^{\gamma T} C e_{Lt}^{})$   \\
$\op{\chi duu}{S}{RL}$ & $\chi \epsilon_{\alpha\beta\gamma}  (d_{Rp}^{\alpha T} C u_{Rr}^{\beta}) (u_{Ls}^{\gamma T} C e_{Lt}^{})$   \\
$\op{\chi dud}{S}{RL}$ & $\chi \epsilon_{\alpha\beta\gamma}  (d_{Rp}^{\alpha T} C u_{Rr}^{\beta}) (d_{Ls}^{\gamma T} C \nu_{Lt}^{})$   \\
$\op{\chi ddu}{S}{RL}$ & $\chi \epsilon_{\alpha\beta\gamma}  (d_{Rp}^{\alpha T} C d_{Rr}^{\beta}) (u_{Ls}^{\gamma T} C \nu_{Lt}^{})$   \\
$\op{\chi duu}{S}{RR}$  & $\chi \epsilon_{\alpha\beta\gamma}  (d_{Rp}^{\alpha T} C u_{Rr}^{\beta}) (u_{Rs}^{\gamma T} C e_{Rt}^{})$  \\
            \hline
        \end{tabular}\\\vspace{5pt}
        \begin{tabular}{c|c}
            \hline
\multicolumn{2}{c}{$\boldsymbol{\Delta B=-\Delta L = 1+ \hc}$} \\
\hline
$\op{\chi ddd}{S}{LL}$ & $\chi \epsilon_{\alpha\beta\gamma}  (d_{Lp}^{\alpha T} C d_{Lr}^{\beta}) (\bar e_{Rs}^{} d_{Lt}^\gamma )$  \\
$\op{\chi udd}{S}{LR}$  & $\chi \epsilon_{\alpha\beta\gamma}  (u_{Lp}^{\alpha T} C d_{Lr}^{\beta}) (\bar \nu_{Ls}^{} d_{Rt}^\gamma )$  \\
$\op{\chi ddu}{S}{LR}$ & $\chi \epsilon_{\alpha\beta\gamma}  (d_{Lp}^{\alpha T} C d_{Lr}^{\beta})  (\bar \nu_{Ls}^{} u_{Rt}^\gamma )$  \\
$\op{\chi ddd}{S}{LR}$ & $\chi \epsilon_{\alpha\beta\gamma}  (d_{Lp}^{\alpha T} C d_{Lr}^{\beta}) (\bar e_{Ls}^{} d_{Rt}^\gamma )$ \\
$\op{\chi ddd}{S}{RL}$  & $\chi \epsilon_{\alpha\beta\gamma}  (d_{Rp}^{\alpha T} C d_{Rr}^{\beta}) (\bar e_{Rs}^{} d_{Lt}^\gamma )$  \\
$\op{\chi udd}{S}{RR}$  & $\chi \epsilon_{\alpha\beta\gamma}  (u_{Rp}^{\alpha T} C d_{Rr}^{\beta}) (\bar \nu_{Ls}^{} d_{Rt}^\gamma )$  \\
$\op{\chi ddd}{S}{RR}$  & $\chi \epsilon_{\alpha\beta\gamma}  (d_{Rp}^{\alpha T} C d_{Rr}^{\beta}) (\bar e_{Ls}^{} d_{Rt}^\gamma )$  \\
            \hline
        \end{tabular}
    \end{minipage}
\end{center}

\section{The Chiral Dilaton Lagrangian}
\label{sec:ChEFT}
Under the chiral scale $\Lambda_{\text{QCD}}\sim 1\text{ GeV}$, the quarks and gluons are confined in the colorless bound states called hadrons, due to the quark condensate. Consequently, the approximate chiral symmetry $SU(3)_L\times SU(3)_R$ is spontaneously broken to the subgroup $SU(3)_V$. 
According to the Goldstone theorem~\cite{Goldstone:1961eq, Goldstone:1962es}, the breaking degrees of freedom generate 8 Nambu-Goldstone bosons (NGBs), which compose the light pseudoscalar meson octets.
The dilaton is not changed during the confinement of the quarks and gluons since it is free of the strong interaction. Thus, we expect it to participate in the interactions of hadrons in various low-energy processes, and we can obtain a chiral theory with dilatons~\cite{Appelquist:2025tol,Appelquist:2022mjb}.

Adopting the CCWZ formalism~\cite{Weinberg:1968de, Coleman:1969sm, Callan:1969sn} developed by Callan, Coleman, Wess, and Zumino, the meson octets can be collected in a unitary matrix,
\begin{equation}
\label{eq:meson}
    u(x) = \exp\left(i\frac{\pi^A(x)\lambda^A}{2f_\pi}\right)\,,
\end{equation}
with
\begin{equation}
    \pi^A(x)\lambda^A = \left(\begin{array}{ccc}
    \pi^0 + \frac{1}{\sqrt{3}}\eta & \sqrt{2}\pi^+ & \sqrt{2}K^+ \\
    \sqrt{2}\pi^- & -\pi^0+\frac{1}{\sqrt{3}}\eta & \sqrt{2}K^0 \\
    \sqrt{2}K^- & \sqrt{2}~\bar{K}^0 & -\frac{2}{\sqrt{3}}\eta
    \end{array}\right)\,,
\end{equation}
where $\lambda^A$ are Gell-Mann matrices. This matrix is covariant as
\begin{equation}
    u(x)\rightarrow V_L u(x)V^\dagger = V u(x)V_R^\dagger\,,
\end{equation}
where $(V_L\,,V_R)\in SU(3)_L\times SU(3)_R$, and $V\in SU(3)_V$. To construct the most general Lagrangian, we define the meson building blocks to be covariant under $SU(3)_V$ solely,
\begin{equation}
    u_\mu = -i\left(u^\dagger \partial_\mu u - u\partial_\mu u^\dagger\right)\,,
\end{equation}
which transforms under $SU(3)_V$ as 
\begin{equation}
    u_\mu \rightarrow V u_\mu V^\dagger\,,\quad V\in SU(3)_V\,.
\end{equation}
The covariant derivative $D_\mu$ is defined as
\begin{equation}
    D_\mu u_\nu = \partial_\mu u_\nu + [\Gamma_\mu,u_\nu]\,,
\end{equation}
where $\Gamma_\mu$ is the chiral connection
\begin{equation}
\Gamma_\mu = \frac{1}{2}(u^\dagger \partial_\mu u + u\partial_\mu u^\dagger)\,.
\end{equation}
The leading-order (LO) Lagrangian is composed of two $u_\mu$,
\begin{equation}
    \mathcal{L} = \frac{f_\pi^2}{4}\text{tr}(u_\mu u^\mu)\,,
\end{equation}
which implies the meson momenta $p$, or the partial derivative of the mesons, is an expansion quantity of the Lagrangian. Thus, the building block $u_\mu$ is of order $p$, and the LO Lagrangian is of order $p^2$.

Because the chiral symmetry is approximate, the mesons are not strictly massless. The mass terms are introduced by external sources,
\begin{equation}
    {\Sigma_\pm} \rightarrow V\Sigma_\pm V^\dagger\,,\quad V\in SU(3)_V\,,
\end{equation}
where
\begin{equation}
    \Sigma_\pm = u^\dagger Mu^\dagger\pm uM^\dagger u\,,
\end{equation}
with $M$ the quark mass matrix, and the subscript $\pm$ implies their parity. 
The pure-meson Lagrangian has been constructed up to $p^8$~\cite{Wess:1971yu,Witten:1983tw,Gasser:1983yg,Gasser:1984gg,Fearing:1994ga,Bijnens:1999hw,Bijnens:1999sh,Ebertshauser:2001nj,Bijnens:2001bb,Li:2024ghg}.

Furthermore, the baryon building blocks could be included by an octet,
\begin{equation}
    B=\left(
    \begin{array}{ccc}
\frac{1}{\sqrt{2}}\Sigma^0+\frac{1}{\sqrt{6}}\Lambda & \Sigma^+ & p \\
\Sigma^- & -\frac{1}{\sqrt{2}}\Sigma^0+\frac{1}{\sqrt{6}}\Lambda & n \\
\Xi^- & \Xi^0 & -\frac{2}{\sqrt{6}}\Lambda \\
\end{array}
    \right)\,,
\end{equation}
which is covariant under $SU(3)_V$, $B\rightarrow VBV^\dagger$.
It is crucial that the baryon masses $m_B$ are comparable to the scale $\lqcd$, and do not vanish under the chiral limit, thus, the power-counting scheme of the baryons is different from that of mesons. The momenta of the baryons are not small quantities because of the on-shell condition,
\begin{equation}
    p^2 = m_B^2 \sim \lqcd^2\,,
\end{equation}
thus, the derivatives of the baryons are powerless. One way to manage the powerless derivatives is the heavy baryon projection (HBP) formalism~\cite{Jenkins:1990jv, Ecker:1995rk, Fettes:2000gb, Kobach:2018pie}. The HBP formalism adopts a time-like vector $v^\mu$ to project out the large component of the momenta $(v\cdot p)v^\mu$ and the small component $v^\mu_T=v^\mu-(v\cdot p)v^\mu$. Consequently, the baryon $B$ can also be projected into different components,
\begin{equation}
    B= \mathcal{B} + \mathcal{H}\,,
\end{equation}
where $\mathcal{B}=P_v^+B$, and $\mathcal{H}=P_v^-B$, with the projectors
\begin{equation}
    P_v^\pm = \frac{1\pm v\cdot\gamma}{2}\,.
\end{equation}
Preserving the component $\mathcal{B}$ and integrating out the 
component $\mathcal{H}$, we obtain an expansion over a small quantity $\frac{1}{v\cdot p}$, which is just the baryon mass $m_B$ if we adopt the rest frame and set $v=(1,0,0,0)$.
Consequently, the Lagrangian in the HBP formalism can be regarded as a simultaneous expansion by both $\lqcd$ and $m_B$. Thus, we define the power of an operator with baryons as the power of both the meson momenta $p$ and the mass inverse $1/m_B$ in the HBP formalism. The LO meson-baryon Lagrangian is 
\begin{equation}
    \mathcal{L}_{\pi B}^{(1)} = \text{tr}(\overline{B}i\gamma^\mu D_\mu B) - \frac{D}{2} \text{tr}(\overline{B}\gamma^\mu\gamma_5\{u_\mu,B\})  - \frac{F}{2} \text{tr}(\overline{B}\gamma^\mu\gamma_5[u_\mu,B]) - \text{tr}(\overline{B}m_B B)\,,
\end{equation}
where the subscript implies it is of power $p$. The meson-baryon operators up to $p^5$ have been constructed.~\cite{Krause:1990xc,Ecker:1995rk,Fettes:1998ud,Fettes:2000gb,Frink:2004ic,Oller:2006yh,Frink:2006hx,Jiang:2016vax,Song:2024fae}

In addition to the building blocks above, the dilaton $\Phi(x)$ is a scalar field transforming trivially under the $SU(3)_V$ symmetry.
Since it is a Nambu-Goldstone boson, it can be expressed in the exponential form
\begin{equation}
    \Phi = f_\chi\exp(\chi/f_\chi)\,,
\end{equation}
similar to the mesons in Eq.\,\eqref{eq:meson}.
Utilizing these building blocks, the effective action involving dilatons should be scale- invariant, which means the effective operators $\mathcal{O}$ should be scaled as
\begin{equation}
    \mathcal{O}(x) \rightarrow e^{d_\mathcal{O}\lambda} \mathcal{O}(e^\lambda x)\,.
\end{equation}
Because the dilaton transforms under the scale symmetry as
\begin{equation}
    \Phi \rightarrow e^\lambda\Phi\Rightarrow 
    \chi \rightarrow 
    \chi+f_\chi\lambda\,,
\end{equation}
we obtain the LO scale-invariant Lagrangian~\cite{Shamir:2021rav,Golterman:2020tdq},
\begin{align}
\label{eq:la1}
    \mathcal{L}_{\pi,\chi}^{(2)} &= \frac{f_\pi^2}{4}e^{2\chi/f_\chi}\text{tr}(u_\mu u^\mu) + \frac{1}{2}e^{2\chi/f_\chi}\left(\partial_\mu\frac{\chi}{f_\chi}\right) \left(\partial^\mu\frac{\chi}{f_\chi}\right) - \frac{f_\pi^2 B_\pi}{2} e^{y\chi/f_\chi}\text{tr}(\Sigma_+), \\
    \mathcal{L}_{\pi B,\chi}^{(1)} &= \text{tr}(\overline{B}i\gamma^\mu D_\mu B) - e^{\chi/f_\chi}\text{tr}(\overline{B}m_B B)\,,
\end{align}
where $y$ is related to the mass anomalous dimension $\gamma_m$ by
\begin{equation}
    y=3-\gamma_m\,.
\end{equation}
The factor $e^{\chi/f_\chi}$ is called the dimension compensator, whose powers are determined by the scale transformations of other fields. 

Under the chiral limit $m=m_u=m_d$, the dilaton-meson and dilaton nucleon interactions can be obtained via the expansion of Eq.\,\eqref{eq:la1},
\begin{align}
    \mathcal{L}_{\pi,\chi}^{(2)} &\supset \frac{\chi}{f_\chi}(\partial_\mu\pi^A\partial^\mu\pi^A) - \frac{m B_\pi y}{2}\frac{\chi}{f_\chi}\pi^A\pi^A\,\notag \\
    &= \frac{4\chi}{f_\chi}(\partial_\mu\pi^+\partial^\mu\pi^- + \frac{1}{2}\partial_\mu\pi^0\partial^\mu \pi^0) -\frac{2mB_\pi y}{f_\chi}\chi(\pi^+\pi^-+\frac{1}{2}\pi^0\pi^0)\,, \\
    \mathcal{L}_{\pi B,\chi}^{(1)} &\supset -m_n\frac{\chi}{f_\chi}\overline{n}n   -m_p\frac{\chi}{f_\chi}\overline{p}p\,.
\end{align}

\subsection{Pure-Meson Operators}

In this subsection, we present the next-to-leading (NLO) pure-meson operators, including the dilatons, which are of $p^4$ order. 
In particular, we use $\langle \dots \rangle$ to represent the traces in the $SU(3)_V$ space.

\subsubsection*{$\calo(p^4)$}

\begin{minipage}{0.48\linewidth}
    \begin{equation}
    \notag
        \begin{array}{|c|}
            \hline
            C+P+ \\
            \hline
            \dl{(y-2)}\lra{u_\mu u^\mu}\lra{\Sigma_+} \\
            \dl{(y-2)}\lra{u_\mu u^\mu \Sigma_+} \\
            \dl{(y-2)}\lra{\Sigma_+\Sigma_+} \\
            \dl{(y-2)}\lra{\Sigma_+}\lra{\Sigma_+} \\
            \dl{(y-2)}\lra{\Sigma_-\Sigma_-} \\
            \dl{(y-2)}\lra{\Sigma_-}\lra{\Sigma_-} \\
            \hline
        \end{array}
    \end{equation}
\end{minipage}
\begin{minipage}{0.48\linewidth}
    \begin{equation}
        \notag
        \begin{array}{|c|}
            \hline
            C+P+ \\
            \hline
            \dl{(y-2)}\lra{u_\mu u^\mu}\lra{\Sigma_-} \\
            \dl{(y-2)}\lra{u_\mu u^\mu \Sigma_-} \\
            \dl{(y-2)}\lra{\Sigma_+\Sigma_-} \\
            \dl{(y-2)}\lra{\Sigma_+}\lra{\Sigma_-} \\
            \hline
        \end{array}
    \end{equation}
\end{minipage}

\subsection{Meson-Baryon Operators}

In this subsection, we present the next-to-leading (NLO) meson-baryon operators including the dilatons, which are of $p^2$ order. 

\subsubsection*{$\calo(p^2)$}

\paragraph{$B^2$:} $ $\newline

\begin{minipage}{0.48\linewidth}
    \begin{equation}
    \notag
        \begin{array}{|c|}
            \hline
            C+P+ \\
            \hline
            \dl{(y-3)}\lra{\overline{B}\Sigma_+B} \\
            \dl{(y-3)}\lra{\overline{B}B\Sigma_+} \\
            \dl{(y-3)}\lra{\overline{B}B}\lra{\Sigma_+} \\
            \dl{(y-3)}\lra{\overline{B}\gamma^5\Sigma_-B} \\
            \dl{(y-3)}\lra{\overline{B}B\gamma^5\Sigma_-} \\
            \dl{(y-3)}\lra{\overline{B}\gamma^5B}\lra{\Sigma_-} \\
            \hline
        \end{array}
    \end{equation}
\end{minipage}
\begin{minipage}{0.48\linewidth}
    \begin{equation}
    \notag
        \begin{array}{|c|}
            \hline
            C+P+ \\
            \hline
            \dl{(y-3)}\lra{\overline{B}\Sigma_-B} \\
            \dl{(y-3)}\lra{\overline{B}B\Sigma_-} \\
            \dl{(y-3)}\lra{\overline{B}B}\lra{\Sigma_-} \\
            \dl{(y-3)}\lra{\overline{B}\gamma^5\Sigma_+B} \\
            \dl{(y-3)}\lra{\overline{B}B\gamma^5\Sigma_+} \\
            \dl{(y-3)}\lra{\overline{B}\gamma^5B}\lra{\Sigma_+} \\
            \hline
        \end{array}
    \end{equation}
\end{minipage}

\paragraph{$B^2u^2$:} $ $\newline
\begin{minipage}{0.48\linewidth}
    \begin{equation}
    \notag
        \begin{array}{|c|}
            \hline
            C+P+ \\
            \hline
            \dl{-}\lra{\overline{B}B}\lra{u_\mu u^\mu} \\
            \dl{-}\lra{\overline{B}u_\mu u^\mu B} \\
            \dl{-}\lra{\overline{B}Bu_\mu u^\mu} \\
            \dl{-}\lra{\overline{B}u_\mu}\lra{B u^\mu} \\
            \dl{-}\lra{\overline{B}\sigma^{\mu\nu}B[u_\mu,u_\nu]} \\
            \dl{-}\lra{\overline{B}\sigma^{\mu\nu}[u_\mu,u_\nu]B} \\
            \dl{-}\lra{\overline{B}\sigma^{\mu\nu}u_\mu}\lra{Bu_\nu} \\
            \dl{-2}\lra{\overline{B}\gamma^\mu\lrd^\nu B}\lra{u_\mu u_\nu} \\
            \dl{-2}\lra{\overline{B}\gamma^\mu\lrd^\nu B \{u_\mu u_\nu\}} \\
            \hline
        \end{array}
    \end{equation}
\end{minipage}
\begin{minipage}{0.48\linewidth}
    \begin{equation}
    \notag
        \begin{array}{|c|}
            \hline
            C+P- \\
            \hline
            \dl{-}\lra{\overline{B}\gamma^5 B}\lra{u_\mu u^\mu} \\
            \dl{-}\lra{\overline{B}\gamma^5 u_\mu u^\mu B} \\
            \dl{-}\lra{\overline{B}\gamma^5 Bu_\mu u^\mu} \\
            \dl{-}\lra{\overline{B}\gamma^5 u_\mu}\lra{B u^\mu} \\
            \dl{-}\lra{\overline{B}\gamma^5 \sigma^{\mu\nu}B[u_\mu,u_\nu]} \\
            \dl{-}\lra{\overline{B}\gamma^5 \sigma^{\mu\nu}[u_\mu,u_\nu]B} \\
            \dl{-}\lra{\overline{B}\gamma^5 \sigma^{\mu\nu}u_\mu}\lra{Bu_\nu} \\
            \dl{-2}\lra{\overline{B}\gamma^5\gamma^\mu \lrd^\nu B[u_\mu,u_\nu]} \\
            \dl{-2}\lra{\overline{B}\gamma^5\gamma^\mu \lrd^\nu [u_\mu,u_\nu] B} \\
            \hline
        \end{array}
    \end{equation}
\end{minipage}

\begin{minipage}{0.48\linewidth}
    \begin{equation}
    \notag
        \begin{array}{|c|}
            \hline
            C-P+ \\
            \hline
            \dl{-}\lra{\overline{B}\sigma^{\mu\nu}u_\mu B u_\nu} \\
            \dl{-2}\lra{\overline{B}\gamma^\mu\lrd^\nu B [u_\mu,u_\nu]} \\
            \dl{-2}\lra{\overline{B}\gamma^\mu\lrd^\nu [u_\mu,u_\nu] B} \\
            \hline
        \end{array}
    \end{equation}
\end{minipage}
\begin{minipage}{0.48\linewidth}
    \begin{equation}
    \notag
        \begin{array}{|c|}
            \hline
            C-P- \\
            \hline
            \dl{-}\lra{\overline{B}\gamma^5\sigma^{\mu\nu}u_\mu B u_\nu} \\
            \dl{-2}\lra{\overline{B}\gamma^5\gamma^\mu\lrd^\nu B}\lra{u_\mu u_\nu} \\
            \dl{-2}\lra{\overline{B}\gamma^5\gamma^\mu\lrd^\nu B\{u_\mu u_\nu\}} \\
            \hline
        \end{array}
    \end{equation}
\end{minipage}

\paragraph{$B^2u\chi$:} $ $\newline
\begin{minipage}{0.48\linewidth}
    \begin{equation}
    \notag
        \begin{array}{|c|}
            \hline
            C+P+ \\
            \hline
            \dl{-}\lra{\overline{B}\gamma^5 B u^\mu}D_\mu\chi \\
            \dl{-}\lra{\overline{B}\gamma^5 u^\mu B}D_\mu\chi \\
            \hline
        \end{array}
    \end{equation}
\end{minipage}
\begin{minipage}{0.48\linewidth}
    \begin{equation}
    \notag
        \begin{array}{|c|}
            \hline
            C+P- \\
            \hline
            \dl{-}\lra{\overline{B} B u^\mu}D_\mu\chi \\
            \dl{-}\lra{\overline{B} u^\mu B}D_\mu\chi \\
            \dl{-2}\lra{\overline{B}\gamma^\mu\lrd^\nu B u_\mu}D_\nu \chi \\
            \dl{-2}\lra{\overline{B}\gamma^\mu\lrd^\nu u_\mu B}D_\nu \chi \\
            \hline
        \end{array}
    \end{equation}
\end{minipage}

\begin{minipage}{0.48\linewidth}
    \begin{equation}
    \notag
        \begin{array}{|c|}
            \hline
            C-P+ \\
            \hline
            \dl{-}\lra{\overline{B}\sigma^{\mu\nu} B u_\mu}D_\nu\chi \\
            \dl{-}\lra{\overline{B}\sigma^{\mu\nu} u_\mu B}D_\nu\chi \\
            \dl{-2}\lra{\overline{B}\gamma^5\gamma^\mu\lrd^\nu B u_\mu}D_\nu \chi \\
            \dl{-2}\lra{\overline{B}\gamma^5\gamma^\mu\lrd^\nu u_\mu B}D_\nu \chi \\
            \hline
        \end{array}
    \end{equation}
\end{minipage}
\begin{minipage}{0.48\linewidth}
    \begin{equation}
    \notag
        \begin{array}{|c|}
            \hline
            C-P- \\
            \hline
            \dl{-}\lra{\overline{B}\gamma^5\sigma^{\mu\nu} B u_\mu}D_\nu\chi \\
            \dl{-}\lra{\overline{B}\gamma^5\sigma^{\mu\nu} u_\mu B}D_\nu\chi \\
            \hline
        \end{array}
    \end{equation}
\end{minipage}

\paragraph{$B^2\chi^2$:} $ $\newline
\begin{minipage}{0.48\linewidth}
    \begin{equation}
    \notag
        \begin{array}{|c|}
            \hline
            C+P+ \\
            \hline
            \dl{-}\lra{\overline{B}B}D_\mu\chi D^\mu\chi \\
            \dl{-2}\lra{\overline{B}\gamma^\mu \lrd B}D_\mu\chi D^\mu\chi \\
            \hline
        \end{array}
    \end{equation}
\end{minipage}
\begin{minipage}{0.48\linewidth}
    \begin{equation}
    \notag
        \begin{array}{|c|}
            \hline
            C+P- \\
            \hline
            \dl{-}\lra{\overline{B}\gamma^5 B}D_\mu\chi D^\mu\chi \\
            \hline
        \end{array}
    \end{equation}
\end{minipage}

\begin{minipage}{0.48\linewidth}
    \begin{equation}
    \notag
        \begin{array}{|c|}
            \hline
            C-P- \\
            \hline
            \dl{-2}\lra{\overline{B}\gamma^5\gamma^\mu \lrd^\nu B}D_\mu\chi D_\nu\chi \\
            \hline
        \end{array}
    \end{equation}
\end{minipage}
\vspace{10pt}

In particular, the $\chi$PT operators presented here involve only the lightest mesons and baryons. The heavier mesons such as $B$ can be introduced via the external sources.

These hadronic operators can be matched from the quark operators in the LEFT.
Because of the confinement, such a matching is non-perturbative. 
The traditional matching method is to utilize various external sources implementing the chiral-symmetry-breaking effects.
However, the external source method becomes involved for high-dimensional operators. An alternative method using spurions rather than the external sources has been developed, under which the matching is organized by the naive dimension analysis and can be extended to the high-dimensional matching systematically~\cite{Song:2025snz}. Besides, the non-perturbative effect can be taken into account via quantitative methods such as lattice computation. The systematic matching deserves more investigation in the future.

\section{Phenomenological Analysis}
\label{sec:pheno}

Building upon the EFT framework derived in the previous sections, we observe that the dilaton couples to a broad spectrum of SM fields across a wide range of energy scales, including gauge bosons, heavy quarks, mesons, and nucleons. This diverse interaction structure permits a comprehensive exploration by combining experimental results related to these particles. 
As mentioned before, while exact scale invariance yields a massless dilaton even at the quantum level, a physical mass $m_\chi$ is generated via explicit symmetry breaking, typically through small deformations from nearly marginal operators. Throughout the parameter space considered in this section, the ratio $m_\chi^2/f_\chi^2$ remains sufficiently small, ensuring that the associated $\mathcal{O}(m_\chi^2/f_\chi^2)$ corrections are negligible and the EFT description remains valid.
In this work, we focus on two typical mass regimes where the dilaton behaves either as a particle-like state or as a coherent wave-like field.

For the particle-like regime, we concentrate on masses around the MeV scale. 
Motivated by the stringent constraints on ALPs from semi-invisible meson decays~\cite{Ertas:2020xcc,Gavela:2019wzg,Izaguirre:2016dfi}, we treat the dilaton as an invisible particle and investigate the specific mass window $m_\chi\in[0.1,200]~$MeV. We restrict our analysis to this range for two reasons: for $m_\chi>2\,m_\mu\simeq 200~$MeV, the dilaton lifetime decreases significantly due to the opening of the decay channel $\chi\to\mu^+\mu^-$, and the signal would be reconstructed from visible dimuon final states, which falls outside the scope of this study. 
For $m_\chi<0.1~$MeV, stellar cooling processes (e.g., in the Sun, red giants, white dwarfs) typically provide the dominant constraints~\cite{Dev:2020jkh,Bottaro:2023gep}. 
While a systematic analysis similar to those performed for ALPs~\cite{Hook:2018dlk} could be carried out for these mass regions, we focus here on the interplay between terrestrial experiments and supernova limits. 
Within the chosen mass window, we probe the dilaton via production at the LHC, rare meson decays and cosmological observations, using the dilaton EFTs derived earlier. 
The relevant Lagrangians governing these processes are given by
\begin{equation}\label{equ:pheno}
\begin{aligned}
    \mathcal{L}_{\rm SM}^{\rm \chi}=&-\frac{\beta_{3}}{2 g_3}\frac{\chi}{f_\chi} G_{\mu \nu}^{A} G^{A \mu \nu}-\frac{\beta_e}{2e} \frac{\chi}{f_\chi}2c_{WW} W^+_{\mu\nu}W^{-\mu\nu}+2\frac{\chi}{f_\chi}m_W^2W_\mu^-W^{+\mu}\\
    &-\frac{\beta_e}{2e} \frac{\chi}{f_\chi}\left(c_{ZZ} Z_{\mu\nu}Z^{\mu\nu}+2c_{\gamma Z} F_{\mu\nu}Z^{\mu\nu}
    +c_{\gamma\gamma} F_{\mu\nu}F^{\mu\nu}\right)+\frac{\chi}{f_\chi}m_Z^2Z_\mu Z^\mu\\
    &-\sum_{\psi=u,d,e}\left(1+\gamma_{m_\psi}\right)m_\psi\frac{\chi}{f_\chi}\bar{\psi}\psi, \\
    \mathcal{L}_{\rm LEFT}^{\chi}\supset& c_{\chi d}^{pr}\,\chi\bar{d}_{Lp}d_{Rr},\\
    \mathcal{L}_{\rm ChPT}^{\rm \chi}\supset&-m_n\frac{\chi}{f_\chi}\bar{n}n - m_p\frac{\chi}{f_\chi}\bar{p}p,
\end{aligned}
\end{equation}
where the dilaton VEV satisfies $f_\chi>1~$TeV, since the scale symmetry breaking scale is much larger than $v$, and the coefficients $c_{ij}$ are given explicitly in Eq.\,\eqref{equ:cij_formula}. 
At the LHC, we focus on the channel $pp\to j\chi$, which has the largest production cross section (see Tab.\,\ref{tab:Production_Xsection_LHC}). 
For rare meson decays, we study $B^+\to K^+\chi$ and $K^+\to\pi^+\chi$, and derive constraints from Belle II and NA62 measurements. 
Finally, we examine the astrophysical impact of dilaton production on the late-time neutrino signal from SN1987A. 
The resulting sensitivities for the conventional dilaton are presented in Fig.\,\ref{fig:constraint}.

In contrast, for the wave-like ultralight regime, we focus on the mass range $m_\chi<10^{-10}~$eV. Above this threshold, constraints are typically dominated by fifth-force searches~\cite{Berge:2017ovy,Hees:2018fpg} sensitive to long-range dilaton exchange, which manifests as violations of the WEP or deviations from the gravitational inverse-square law. In the ultralight regime considered here, the primary observable is the oscillation of the fine-structure constant driven by the dilaton-photon coupling. Constraints on the dilaton VEV, derived from precision measurements of these oscillations using atomic clocks and atom interferometers, are summarized in Fig.\,\ref{fig:atomic_constraint}.

\subsection{Survival probability of the dilaton}
For the mass regimes in which the dilaton behaves as a conventional particle, it can decay into both photon pairs and electron pairs. 
It is therefore necessary to account for the effect of dilaton decays in our analysis. 
Within the Lagrangian given in Eq.\,\eqref{equ:pheno}, the partial decay widths for $\chi\to\gamma\gamma$ and $\chi\to e^+e^-$ are, at tree-level,
\begin{equation}
\begin{aligned}\label{equ:dilaton_decay}
\Gamma(\chi\rightarrow\gamma\gamma)&=\left(\frac{\beta_e}{2e}c_{\gamma\gamma}\right)^2\frac{m_\chi^3}{4\pi f_\chi^2},\\
    \Gamma(\chi\rightarrow e^+e^-)&=\frac{(\gamma_{m}+1)^2m_e^2m_\chi}{8\pi f_\chi^2}\left(1-\frac{4m_e^2}{m_\chi^2}\right)^{\frac{3}{2}}.
\end{aligned}
\end{equation}
Using these channels, the decay length of the dilaton in the laboratory frame is 
\begin{equation}
    d=\gamma c\tau_{\chi}=\frac{\hbar}{\Gamma_{\rm tot}(\chi)}\frac{E_\chi}{m_\chi}c,
\end{equation}
where $\Gamma_{\rm tot}(\chi)$ and $\tau_\chi$ are the total decay width and proper lifetime of the dilaton, $\gamma=E_\chi/m_\chi$ is its Lorentz boost factor, and $E_\chi$ denotes its energy. 
If the dilaton VEV $f_\chi$ or its energy $E_\chi$ is sufficiently small, the produced dilaton may decay into visible $e^+e^-$ or $\gamma\gamma$ pairs inside the detector. 
To incorporate this effect, we define a survival factor for each of the experiments,
\begin{equation}\label{equ:survivingrate}
    f_{\rm sur,exp}=e^{-L_{\rm exp}/d},
\end{equation}
where $L_{\rm exp}$ is the relevant detector size. 
This factor represents the probability that the dilaton remains invisible (i.e., does not decay visibly) while traversing the detector, and it must be included when deriving constraints on the dilaton parameter space.

In the subsequent LHC phenomenology, the produced dilatons are highly boosted. 
Nevertheless, for large dilaton masses or small $f_\chi$, there remains a small probability for decays into $e^+e^-$ and $\gamma\gamma$ inside the detector. 
We therefore include the survival factor in our analysis, adopting a typical dilaton energy $E_{\chi,\,\rm LHC}=120~$GeV\footnote{This choice reflects the requirement imposed in the mono-jet selection of our analysis.} and a detector size $L_{\rm LHC}=10~$m. 
For the search via $B^+\to K^+\chi$ at Belle~II, we use a representative laboratory-frame energy $E_{\chi,\,\rm Belle~II}=3~$GeV and a detector size $L_{\rm Belle~II}=8~$m. 
For the channel $K^+\to \pi^+\chi$ at NA62, we take $E_{\chi,\,\text{NA62}}=30$~GeV and $L_\text{NA62}=150$~m. 
With these experimental parameters, the survival factor can be evaluated for collider and rare-decay measurements, thereby yielding realistic constraints on the dilaton VEV $f_\chi$.

\subsection{Dilaton Production at LHC}

\begin{figure}
\centering
\includegraphics[width=0.75\textwidth]{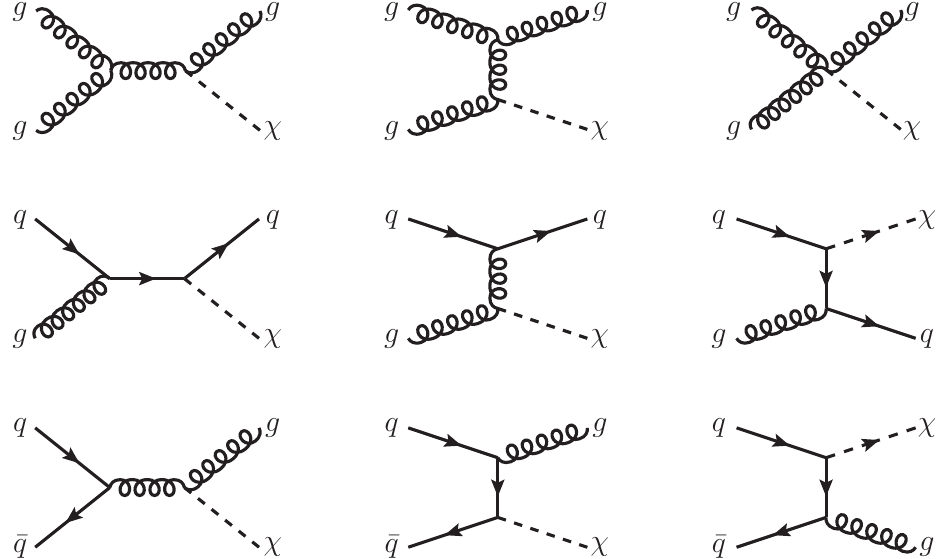}
\caption{Representative Feynman diagrams for the signal process $pp\to j\chi$.
} \label{fig:monojet_signal_FeynmanDiagram}
\end{figure}

The Lagrangian in Eq.\,\eqref{equ:pheno} implies direct couplings between the dilaton and SM elementary particles. 
Production cross sections for the dilaton through various channels are listed in  Tab.\,\ref{tab:Production_Xsection_LHC}. 
\begin{table}
\caption{Production cross sections of the dilaton in different scattering channels at the HL-LHC, where the dilaton VEV and mass are fixed to $f_\chi=10~\mathrm{TeV}$ and $m_\chi=1~\mathrm{MeV}$, respectively. 
The production cross sections for other values of the dilaton VEV can be obtained by rescaling with a factor proportional to $1/f_\chi^2$.  
}\label{tab:Production_Xsection_LHC}
    \centering
    \begin{tabular}{c|c|c|c|c|c}
    \hline
    \hline
    Process &$pp\to t\bar{t}\chi$  &$pp\to j\chi$   &$pp\to W\chi$  &$pp\to Z\chi$  &$pp\to\gamma\chi$ \\
    \hline
    $\sigma$ [pb] &$0.075$   &$7.11$  &$0.58$  &$0.21$  &$4.45\times 10^{-5}$ \\
    \hline
    \hline
    \end{tabular}
\end{table}
In this work we focus on the channel $pp\to j\chi$, which yields the largest cross section. 
Representative Feynman diagrams for these processes are shown in Fig.\,\ref{fig:monojet_signal_FeynmanDiagram}.
Before decaying into $e^+e^-$ or $\gamma\gamma$, the dilaton behaves as an invisible particle at a high-energy. 
Consequently, the $pp\to j\chi$ channel gives rise to a mono-jet signature, which has been extensively studied in previous LHC searches~\cite{CMS:2011esc,CMS:2014jvv,CMS:2021far}. 
The dominant SM backgrounds for the mono-jet final state are
\begin{equation}
\begin{aligned}
pp\to2j,3j,4j,~~pp\to jW^\pm(\to \ell^\pm\nu),~~pp\to jZ(\to\nu\bar{\nu}), \nonumber
\end{aligned}
\end{equation}
where $\ell=e,\mu$. 
For simulation, the dilaton model is implemented in FeynRules~\cite{Alloul:2013bka}. 
Signal and background events are generated with MadGraph~\cite{Alwall:2014hca}, followed by parton showering and hadronization performed with Pythia8~\cite{Sjostrand:2014zea}. 
Detector effects are incorporated using Delphes~\cite{deFavereau:2013fsa}.

To optimize the sensitivity to the dilaton, a set of kinematic selection cuts is applied to suppress SM backgrounds. 
These cuts are summarized in Tab.\,\ref{tab:optimized_dilaton}.
\begin{table}
\caption{Optimized selection cuts for dilaton production via $pp\to j\chi$ at LHC.}
    \label{tab:optimized_dilaton}
    \centering
    \begin{tabular}{c c c}
    \hline\hline
    Cuts &Description &  Values \\
    \hline 
    1   &Mono-jet signature &$\slashed{E}_T>120$~GeV, $0<N_j\leq2$ \\
    \hline
    2   &Leading jet kinematics  &$P^T_{j_1}>60$~GeV, $|\eta_{j_1}|<2.5$, \\ 
    \hline
    3   &Sub-leading jet kinematics   &$\Delta\phi(j_1,j_2)<2.0$, \\
    \hline
    4   &Veto isolated lepton & $N_\ell=0.$\\
    \hline
    \hline
    \end{tabular}
\end{table}
Here, $\slashed{E}_T$ denotes the transverse missing energy, and $N_j$ is the number of jets with $P^T_j > 20$~GeV and $|\eta_j|<5.0$. 
$P^T_i$ and $\eta_i$ denote the transverse momentum and pseudo-rapidity of particle $i$, respectively. 
The azimuthal separation between the leading and subleading jet is $\Delta\phi(j_1,j_2)=|\phi_{j_1}-\phi_{j_2}|$, and $N_\ell$ counts charged lepton satisfying $P^T_\ell>10~$GeV and $|\eta_\ell|<2.5$. 
A large missing transverse energy requirement is essential to define the mono-jet signature. 
To retain signal efficiency, the presence of a second jet is allowed. 
Cuts on the leading jet ($P^T_{j_1}>60$ GeV, $|\eta_{j_1}|<2.5$) and on the azimuthal separation between the two jets ( $\Delta\phi(j_1,j_2)<2.0$) significantly suppress the multi-jet QCD background. 
Finally, events containing an isolated charged lepton are vetoed ($N_\ell=0$) to reduce the background from $jW^\pm$ production.
The cutflow for the signal and the dominant backgrounds is presented in 
Tab.\,\ref{tab:Cutflow}. 
\begin{table}
\caption{Cutflow for the signal and SM backgrounds. The dilaton benchmark point is taken as $m_{\chi}=1$~MeV,  $f_\chi=10$~TeV.}
    \label{tab:Cutflow}
\resizebox{1.0\columnwidth}{!}{
    \centering
    \begin{tabular}{c|c|c|c|c|c|c}
    \hline
    \hline
    \multirow{2}{*}{Cuts} &\multicolumn{6}{c}{Cross section for signal and background [pb] } \\
    \cline{2-7}
    &$pp\to j\chi$ &$pp\to 2j$  &$pp\to 3j$  &$pp\to4j$  &$pp\to jW^\pm(\to\ell^\pm\nu)$ &$pp\to jZ(\to\nu\bar{\nu})$\\
    \hline
    Cut $1$ &$0.078$  &$13308$  &988.85   &108.18  &$54.99$  &$52.43$ \\
    \hline
    Cut $2$ &$0.072$  &$522$  &466.96   &42.07  &$51.70$  &$49.77$ \\
    \hline
    Cut $3$ &$0.059$  &$102$  &47.70   &15.14  &$40.96$  &$41.13$ \\
    \hline
    Cut $4$ &$0.059$  &$102$  &47.70   &15.14  &$21.48$  &$41.12$ \\
    \hline
    \hline
    \end{tabular}
    }
\end{table}

To quantify the sensitivity to the dilaton signal, we define the $90\%$ confidence level (C.L.) exclusion significance based on a likelihood analysis~\cite{Cowan:2010js} as
\begin{equation}
\begin{aligned}
\sigma_{\text{exc}} &= \sqrt{-2\,\ln \left( \frac{L(S+B| B)}{ L (B|B)} \right) } \geq 1.65, 
\end{aligned} \label{equ:likelihood_exclusion}
\end{equation}
where $L$ denotes the Poisson likelihood function.
This criterion corresponds to a one-sided 90\% C.L. exclusion under the asymptotic approximation.
Here, $B$ represents the expected number of background events at the HL-LHC with an integrated luminosity of $\mathcal{L}=3~\mathrm{ab}^{-1}$, and $S$ denotes the number of signal events, which scales as $1/f_\chi^2$. 
The Poisson likelihood function is defined as 
\begin{equation}
    L(X | Y) = \frac{X^{Y}}{Y!} \, e^{-X}. 
\end{equation}
Within this framework, the resulting constraints on $f_\chi$ as a function of the dilaton mass in the range $m_\chi\in[0.1,200]~\mathrm{MeV}$ are shown as the purple shaded region in Fig.\,\ref{fig:constraint}. 
We find that the HL-LHC is sensitive to a dilaton vacuum expectation value up to approximately $20~\mathrm{TeV}$ over the considered mass range.

\subsection{Semi-invisible decays of $B$ and $K$ meson}

In addition to direct collider searches, the dilaton can be probed through rare semi-invisible decays of $B$ and $K$ mesons, provided its mass is below the relevant kinematic thresholds: $m_\chi<m_{B^+}-m_{K^+}$ or $m_\chi<m_{K^+}-m_{\pi^+}$.

\begin{figure}
    \centering
    \includegraphics[width=0.8\linewidth]{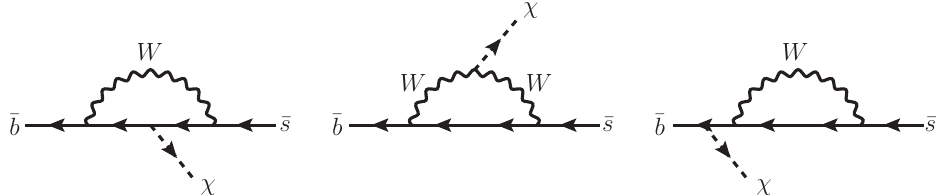}
    \caption{Representative one-loop Feynman diagrams for $B^+\rightarrow K^+\chi$ channel.}
    \label{fig:FCNC-dilaton}
\end{figure}

For the $B$ meson system, the flavor-changing neutral current transition $\bar{b}\to \bar{s}\chi$ arises at the one-loop level. 
Representative Feynman diagrams are shown in Fig.\,\ref{fig:FCNC-dilaton}. 
These contributions are encapsulated in the effective Lagrangian
\begin{equation}\label{equ:Lag_bs}
    \mathcal{L}_{\chi bs}=\frac{(\gamma_{m_b}+1)m_b}{f_\chi}  \frac{3 \sqrt{2} G_Fm_{t}^{2} V_{t s}^{*} V_{t b}}{16 \pi^{2}}  \chi \bar{b}_{R} s_{L}+\text { h.c. },
\end{equation}
where $V_{ts,tb}$ are Cabibbo-Kobayashi-Maskawa (CKM) matrix elements.
The contribution from the operators $\chi\bar{b}_Ls_R$ is neglected, as it is suppressed by a factor of $m_s/m_b\sim0.02$ compared to Eq.\,\eqref{equ:Lag_bs}. 
Under the effective Lagrangian of dilaton, the branching ratio for $B^+\to K^+\chi$ is
\begin{equation}
\begin{aligned}\label{equ:BR_BtoK}
\text{Br}(B^+ \rightarrow K^+ \chi)=&\frac{1}{\Gamma_{B^+}}\left[\frac{(\gamma_{m_b}+1)m_b}{f_\chi}  \frac{3 \sqrt{2} G_Fm_{t}^{2} V_{t s}^{*} V_{t b}}{16 \pi^{2}}\right]^{2} \frac{\lambda_{B^+\rightarrow K^+\chi}^{1/2}}{16 \pi m_{B^+}^{3}} \\
& \left.\times\left|\langle K^+| \bar{s}_{L} b_{R}\right| B^+\right\rangle\left.\right|^{2} \Theta\left(m_{B^+}-m_{K^+}-m_\chi\right), 
\end{aligned}
\end{equation}
where $\lambda_{X\to YZ}=\left[m_X^2-(m_Y+m_Z)^2\right]\left[ m_X^2-(m_Y-m_Z)^2 \right]$, $\Gamma_{B^+}$ is the total decay width of $B^+$ meson, and the hadronic matrix element is given by~\cite{Gubernari:2018wyi}
\begin{equation}
    \langle K^+| \bar{s}_{L} b_{R}|B^+\rangle=\frac{m_{B^+}^{2}-m_{K^+}^{2}}{2\left(m_{b}-m_{s}\right)} f_{0}^{BK}\left(m_{\chi}^{2}\right).
\end{equation}
Here $f_0^{BK}$ is the form factor for the $B^+\to K^+$ transition. 
We adopt the parameterization from Ref.~\cite{Bharucha:2015bzk},
\begin{equation}
    f_{0}^{BK}\left(q^{2}\right)=\frac{1}{1-q^{2} / m_{R}^{2}} \sum_{k=0,1,2} a_{k}\left[z\left(q^{2}\right)-z(0)\right]^{k},
    \label{formfactor}
\end{equation}
with a resonance mass $m_R=5.711~$GeV. 
The conformal mapping variable is
\begin{equation}
    z(q^2)=\frac{\sqrt{t_{+}-q^2}-\sqrt{t_{+}-t_{0}}}{\sqrt{t_{+}-q^2}+\sqrt{t_{+}-t_{0}}} ,
\end{equation}
where $t_{0}=t_{+}\left(1-\sqrt{1-t_{-} / t_{+}}\right)$ and $t_{ \pm}=\left(m_{B^+} \pm m_{K^+}\right)^{2}$. 
The coefficients $a_k$ fitted to a combination of lattice QCD results from the HPQCD collaboration~\cite{Parrott:2022rgu} and light-cone sum rule (LCSR) calculation~\cite{Gubernari:2018wyi}, are $a_0=0.3233(67),\quad a_1=0.214(57),\quad a_2=-0.12(13)$~\cite{Grunwald:2023nli}.

The dominant SM background for the $B^+\to K^+\chi$ search is the decay $B^+\to K^+\nu\bar{\nu}$. 
Recently, the Belle~II collaboration reported a branching ratio measurement of $\text{Br}(B^+\to K^+\nu\bar{\nu})_\text{exp}=[2.3\pm0.5(\text{stat})^{+0.5}_{-0.4}(\text{syst})]\times10^{-5}$~\cite{Belle-II:2023esi}, which exceeds the SM prediction $\text{Br}(B^+\to K^+\nu\bar{\nu})_\text{SM}=[0.497\pm0.037]\times10^{-5}$~\cite{Parrott:2022zte} by approximately $2.7\sigma$. 
Interpreting this excess as a potential contribution from new physics, the $90\%$ C.L. allowed range for the non-SM branching ratio is $\text{Br}(B^+\to K^+ + \text{inv})_\text{NP}=(1.8^{+1.2}_{-1.1})\times10^{-5}$. 
If this excess is attributed to the dilaton, the branching ratio in Eq.\,\eqref{equ:BR_BtoK} is required to satisfy
\begin{equation}
    0.7\times 10^{-5}<{\rm Br}(B^+\to K^+\chi)\,f_\text{sur,Belle \uppercase\expandafter{\romannumeral2}} < 3.0\times 10^{-5}.
\end{equation}
This requirement corresponds to the allowed parameter space enclosed by the blue dot-dashed lines in Fig.\,\ref{fig:constraint}, which restricts $f_\chi$ to a range between approximately $35~\mathrm{TeV}$ and $70~\mathrm{TeV}$. 

On the other hand, the excess may also be explained by statistical or systematic fluctuations. 
In this case, we derive a $90\%$ C.L. constraint on the dilaton VEV using the experimental uncertainty at Belle~II, which leads to 
\begin{equation}
    {\rm Br}(B^+\to K^+\chi)\,f_\text{sur,Belle \uppercase\expandafter{\romannumeral2}}<1.2\times 10^{-5}. 
\end{equation}
The corresponding exclusion region in the $(m_\chi, f_\chi)$ plane is shown as the blue shaded region in Fig.\,\ref{fig:constraint}. 
We find that the resulting constraint on $f_\chi$ can reach up to approximately $55~\mathrm{TeV}$, which is stronger than that from direct collider searches over the entire considered mass range $m_\chi\in[0.1,200]~\mathrm{MeV}$. 

Analogously, constraints can be derived from the rare decay $K^+\to\pi^+\chi$. 
The relevant effective Lagrangian is obtained from Eq.\,\eqref{equ:Lag_bs} by replacing $\bar{b}\to\bar{s}$ and final $\bar{s}\to\bar{d}$. 
The branching ratio is
\begin{equation}
\begin{aligned}
\text{Br}({K^+ \rightarrow \pi^+ \chi})=&\frac{1}{\Gamma_{K^+}}\left[\frac{(\gamma_{m_s}+1)m_s}{f_\chi}  \frac{3 \sqrt{2} G_Fm_{t}^{2} V_{t d}^{*} V_{t s}}{16 \pi^{2}}\right]^{2} \frac{\lambda_{K^+\rightarrow\pi^+\chi}^{1/2}}{16 \pi m_{K^+}^{3}} \\
& \left.\times\left|\langle \pi^+| \bar{d}_{L} s_{R}\right| K^+\right\rangle\left.\right|^{2} \Theta\left(m_{K^+}-m_{\pi^+}-m_\chi\right),
\end{aligned}           
\end{equation}
where the hadronic matrix element is~\cite{Abdughani:2023dlr,Gubernari:2018wyi}
\begin{equation}
    \langle \pi^+| \bar{d}_{L} s_{R}|K^+\rangle=\frac{m_{K^+}^{2}-m_{\pi^+}^{2}}{2\left(m_{s}-m_{d}\right)} f_{0}^{K\pi}\left(m_{\chi}^{2}\right).
\end{equation}
The form factor $f_0^{K\pi}(q^2)$ is very close to unity for $0<q^2\lesssim(m_K-m_\pi)^2$~\cite{Boyle:2010bh,Carrasco:2016kpy}. 
The primary SM background in this channel is $K^+\to\pi^+\nu\bar{\nu}$. 
The latest measurements from NA62 have already measured the differential distribution of branching ratios versus the mass of invisible new particle at $90\%$ C.L. (as shown in Fig.\,8 in Ref.~\cite{NA62:2021zjw}). 
The constraint on the dilaton is therefore given by
\begin{equation}
    \text{Br}({K^+ \rightarrow \pi^+ \chi})f_\text{sur,NA62}<\mathrm{Br}(K^+ \to \pi^++inv)_{\rm NA62}.
\end{equation}
The corresponding excluded region is also presented in Fig.\,\ref{fig:constraint}. 
The sensitivity from NA62 reaches $f_\chi\sim2000~$TeV, approximately one order of magnitude stronger than the limits from Belle~II and the HL-LHC. 
However, a weaker exclusion is observed for $m_\chi\in[100,150]~$MeV due to the large irreducible background from $K^+\to\pi^+\pi^0$ decay in that mass region. 
For this specific window, the Belle-II and HL-LHC searches provide complementary probes of the dilaton parameter space.

\subsection{Dilaton production in SN1987A}
For dilaton masses below a few hundred MeV, on-shell production can occur within the core of compact astrophysical objects. 
Inspired by extensive studies on production and subsequent signatures of ALPs in various high-energy astrophysical environments~\cite{Caputo:2022mah,Fiorillo:2025yzf,Diamond:2023cto,Diamond:2023scc}, the production and decay processes of dilatons might significantly influence the evolution and observational signatures of these objects~\cite{Dev:2020eam,Balaji:2022noj}.
In this work, we primarily utilize the observed late-time neutrino signal from SN1987A~\cite{Kamiokande-II:1987idp} to constrain the dilaton's couplings to matter.
The standard cooling bound is obtained by requiring that the luminosity carried away by dilatons does not exceed the benchmark neutrino luminosity $\mathcal{L}_{\rm SN}\sim 10^{53}\,{\rm erg/s}$ inferred from the SN1987A neutrino data.

Following Refs.~\cite{Girmohanta:2023tdr, Krnjaic:2015mbs,Ishizuka:1989ts}, the energy-loss rate per unit volume due to dilaton emission from the supernova core is estimated as 
\begin{equation}\label{equ:emission_rate}
    Q_\chi\approx \left(\frac{m_N}{f_\chi}\right)^2\frac{11}{(15\pi)^3} \left( \frac{T_{SN}}{m_\pi} \right)^4p_F^5\,G_{\chi}\left( \frac{m_\pi}{p_F} \right)\xi(T_{SN},m_\chi),
\end{equation}
where $m_N/f_\chi$ denotes the effective dilaton-nucleon coupling, $T_{SN}$ is the core temperature, $m_\pi$ is the pion mass, and $p_F$ is the nucleon Fermi momentum. 
The dimensionless function $G_\chi$ is given by
\begin{equation}
\begin{aligned}\label{equ:Gchi(u)}
    G_\chi(u)=& 1-\frac{5}{2}u^2-\frac{35}{32}u^4+\frac{5}{64}(28u^3+5u^5)\,{\rm arctan}\left( \frac{2}{u} \right) \\
        &+\frac{5}{64}\frac{\sqrt{2}u^6}{\sqrt{2+u^2}}\,{\rm arctan}\left( \frac{2\sqrt{2(u^2+2)}}{u^2} \right).
\end{aligned}
\end{equation}
To approximately account for finite-mass effects, we introduce a correction factor $\xi(T,m_\chi)$ following Ref.~\cite{Krnjaic:2015mbs}:
\begin{equation}\label{equ:Xi_chi}
\xi(T,m_\chi)\equiv\frac{\int_{m_\chi}^\infty dx\frac{x\sqrt{x^2-m^2_\chi}}{e^{x/T}-1}}{\int_0^\infty dx\frac{x^2}{e^{x/T}-1}}. 
\end{equation}
This factor captures the leading phase-space suppression associated with the relativistic dispersion relation $E=\sqrt{p^2+m_\chi^2}$ for on-shell dilaton emission, effectively replacing the massless Bose--Einstein integral $\int_0^\infty dx\,x^2/(e^{x/T}-1)$ with its massive counterpart $\int_{m_\chi}^\infty dx\,x\sqrt{x^2-m_\chi^2}/(e^{x/T}-1)$. 
The approximation is valid when $m_\chi\lesssim \mathcal{O}({\rm few})\times T_{SN}$ and the dilaton remains in the free-streaming regime without significant trapping effects. 
In this regime, $\xi$ interpolates between the relativistic limit ($\xi\to1$ for $m_\chi\ll T_{SN}$) and the Boltzmann-suppressed regime ($m_\chi\gg T_{SN}$). 
For the benchmark parameters adopted in our analysis (See Table~\ref{tab:Input_Para}), this correction remains within the overall order-of-magnitude uncertainty of the SN1987A cooling bound. 
A fully numerical treatment including detailed production and trapping effects would be required for higher precision, which is beyond the scope of the present estimation. 

After being produced in the core, the dilaton may decay or be reabsorbed while propagating through the supernova medium. 
This effect is approximately incorporated via an escape probability
\begin{equation}\label{equ:escape_probability}
    P_{\rm esc.}\simeq e^{-R_{SN}/(\gamma c\tau_\chi)}e^{-R_{SN}Q_\chi/\rho_\chi},
\end{equation}
where $R_{SN}$ is the core radius, and $\gamma$ and $\tau_\chi$ denote the boost factor and lifetime (including the decay channels in Eq.\,\eqref{equ:dilaton_decay}) of the dilaton, and $\rho_\chi$ is the dilaton energy density estimated in the equilibrium limit at the supernova core temperature $T_{SN}$. 

\begin{figure}
    \centering
\includegraphics[width=\linewidth]{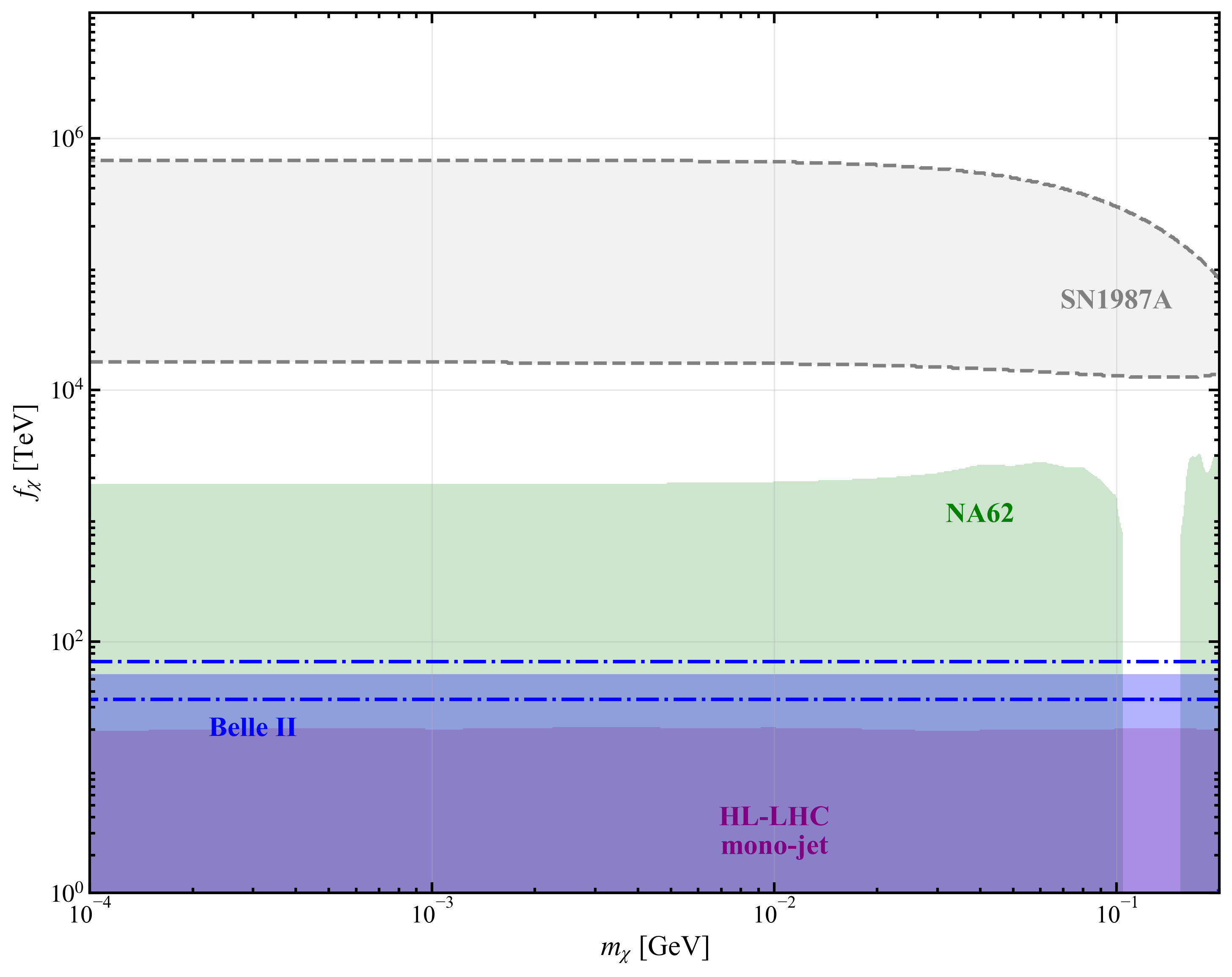}
\caption{Constraints on the dilaton VEV $f_\chi$. 
The purple shaded region denotes the exclusion from HL-LHC. 
The region between the two blue dot-dashed lines corresponds to the $90\%$ C.L. allowed region for the dilaton VEV when the excess observed at Belle\,II~\cite{Belle-II:2023esi} is interpreted as a signal of new physics, while blue shaded region denotes the exclusion obtained under the assumption that the excess arises from statistical or systematic fluctuations. 
The green shaded region shows the exclusion from the NA62 collaboration~\cite{NA62:2021zjw}. 
The gray dashed contour indicates the approximate exclusion derived from SN1987A~\cite{Kamiokande-II:1987idp}, where the input parameters used in the estimation are summarized in Tab.\,\ref{tab:Input_Para}.}
\label{fig:constraint}
\end{figure}

Finally, the constraint from SN1987A is obtained by requiring the total energy-loss rate to satisfy
\begin{equation}\label{equ:boundary_SN198A}
    P_{\rm esc.}\, Q_\chi \,V_{SN}\lesssim\,\mathcal{L}_{\rm SN},
\end{equation}
where $V_{SN}=(4\pi/3)R_{SN}^3$ is the core volume. 
The corresponding constraints on the dilaton VEV are shown as the gray dashed region in Fig.\,\ref{fig:constraint}. 
For $m_\chi\in[0.1,200]~\mathrm{MeV}$, we obtain an exclusion of $f_\chi\sim\mathcal{O}(10^4\text{--}10^6)~\mathrm{TeV}$, consistent with existing supernova bounds on dilaton~\cite{Girmohanta:2023tdr}.

\begin{table}
\caption{Typical values of input parameters used in the estimation of SN1987A constraints, following Ref.~\cite{Krnjaic:2015mbs}. }
\label{tab:Input_Para}
\vspace{0.2cm}
\resizebox{1.0\columnwidth}{!}{
    \centering
    \begin{tabular}{c|c|c|c|c|c|c}
    \hline
    \hline
    Parameter &$p_F$  &$T_{SN}$   &$m_\pi$  &$m_N$  &$R_{SN}$ &$\gamma$ \\
    \hline
    \multirow{2}{*}{Value} &\multirow{2}{*}{$200$~MeV}   &\multirow{2}{*}{$30~$MeV}  &\multirow{2}{*}{$140$~MeV}  &\multirow{2}{*}{$940$~MeV}  &\multirow{2}{*}{$10~$km} &$3\,T_{SN}/m_\chi$ (for $m_\chi<3\,T_{SN}$) \\
    & & & & & & $1$ (for $m_\chi\geq3\,T_{SN}$) \\ 
    \hline
    \hline
    \end{tabular}
    }
\end{table}

\subsection{Atomic clocks and interferometers}
\label{sec:atomic}
\begin{figure}
    \centering
    \includegraphics[width=\linewidth]{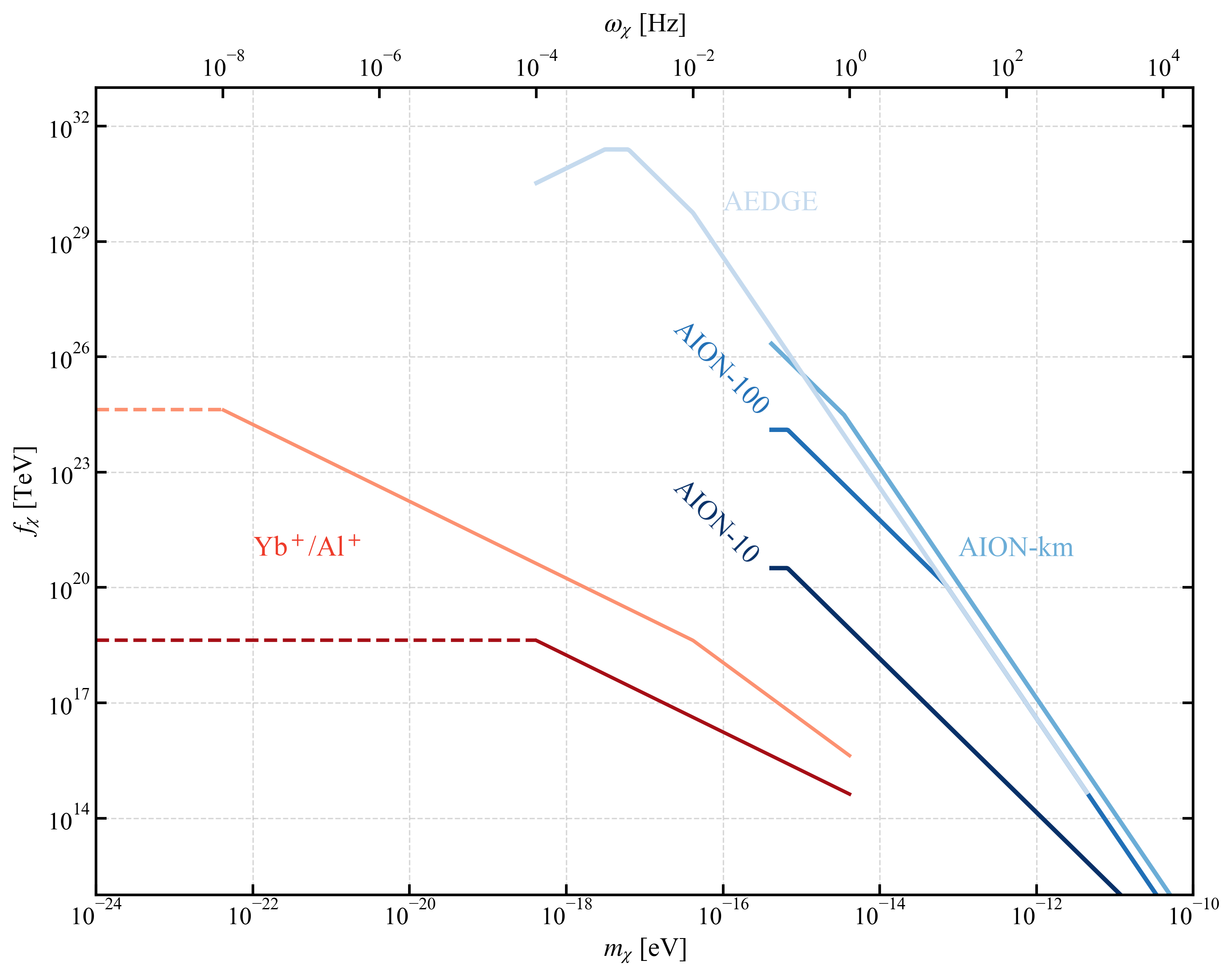}
    \caption{Sensitivity projections for the dilaton VEV $f_\chi$ as a function of dilaton mass $m_\chi$. Red curves depict the SNR = 1 sensitivity envelopes of the Yb$^+$/Al$^+$ clock comparison experiment, for $\tau_\text{int}=10^4,10^8$~s (deep red, light coral). Also depicted are sensitivities of different AION stages.}
    \label{fig:atomic_constraint}
\end{figure}
In this section, we derive constraints on the dilaton as a candidate for ultralight dark matter (ULDM)~\cite{Kimball:2023vxk}. In the regime $m_\chi\ll1$~eV, the high galactic occupation number allows the DM to be treated as a classical coherent field. Assuming the dilaton saturates the local DM density, $\rho_\mathrm{DM} \approx 0.3\,\mathrm{GeV/cm^3}$, the field evolves as
 \begin{equation}
    \chi(\mathbf{x},t)=\frac{\sqrt{2\rho_\text{DM}}}{m_\chi}\cos[m_\chi(t-\mathbf{v}_\chi\cdot\mathbf{x})+\cdots],
\end{equation}
where $\mathbf{v}_\chi$ is the local DM velocity. The velocity dispersion $v_\text{vir}\sim10^{-3}c$ implies a coherent time $\tau_\text{coh}\approx2\pi/(m_\chi v_\text{vir}^2)$, during which the field behaves as a monochromatic wave.

Couplings between the dilaton and SM fields, as introduced in Eq.\,\eqref{equ:QCD+QED+dilaton}, induce temporal oscillations in fundamental constants. Specifically, the photon coupling leads to variations in the fine-structure constant $\alpha$:
\begin{equation}
    \frac{\Delta\alpha(t)}{\alpha}\approx-\frac{2\beta_e}{e}c_{\gamma\gamma}\frac{\chi}{f_\chi}.
\end{equation}
Analogous variations occur in fermion masses and the QCD scale, thereby modulating atomic transition energies. These energies depend on $\alpha$ and the proton-to-electron mass ratio $\mu = m_p/m_e$, scaling as:
\begin{equation}
    f_A(t)\propto\alpha^{\xi_A+2} \mu^{\zeta_A}.
\end{equation}
Here, the explicit $\alpha^2$ dependence arises from the Rydberg constant ($R_\infty \propto \alpha^2$), while $\xi_A$ parametrizes the sensitivity of relativistic corrections specific to the transition $A$~\cite{Dzuba:1998au}. The exponent $\zeta_A$ characterizes the dependence on the mass ratio, taking values of $\zeta_A \simeq 1$ for hyperfine transitions and $\zeta_A \simeq 0$ for optical transitions.

Atomic precision experiments~\cite{Antypas:2022asj,Safronova:2017xyt,Derevianko:2022lmn} provide the most stringent constraints on such dilatons, especially via searches for oscillating $\alpha$. Atomic clock comparisons~\cite{Arvanitaki:2014faa,Filzinger:2023zrs} exploit transitions with different $\alpha$ sensitivities. The fractional variation in the frequency ratio of two optical transitions, $\nu_A$ and $\nu_B$, is given by 
\begin{equation}
\frac{\Delta(\nu_A/\nu_B)}{\nu_A/\nu_B} = (\xi_A-\xi_B) \frac{\Delta\alpha}{\alpha}\,,
\end{equation}
Fig.\,\ref{fig:atomic_constraint} shows the projected unit signal-to-noise ratio (SNR = 1) sensitivity to the VEV $f_\chi$, assuming a sampling rate of $1$ Hz. We consider E3 transitions in $^{171}$Yb$^+$ compared to $^{27}$Al$^+$, for integration times of $\tau_\text{int}=10^4$~s and $\tau_\text{int}=10^8$~s. The sensitivity scales with the number of coherent measurements, enhancing stability by a factor $\beta=(\min\{\tau_\text{int},\tau_\text{coh}\}/\Delta\tau)^{1/2}$, where $\Delta\tau$ is the sampling interval. The detection strategy depends on the DM mass relative to the integration time. We require the dilaton frequency $\omega_\chi=m_\chi/2\pi$ below the sampling rate for an isolated spectral peak. For $2\pi/m_\chi>\tau_\text{int}$, the dilaton field evolves slowly, manifesting as a linear frequency drift indistinguishable from systematic noise. In this regime, the bounds become mass-independent, represented by the flat extension in the sensitivity curves.

Complementary to co-located clocks, atom interferometry~\cite{Cronin:2009zz} exploits spatially separated interferometers acting on a single atomic species referenced to a common laser. The ULDM-induced oscillation in the transition energy, $\Delta f_A/f_A=\xi_A(\Delta\alpha/\alpha)$, generates a differential phase shift. For an interferometer with baseline $L$, interrogation time $2T$ and $n$ large momentum transfer pulses, the signal phase amplitude is~\cite{Arvanitaki:2016fyj}
\begin{equation}
    \bar{\Phi}_{s}=8 \frac{\Delta f_{A}}{m_{\chi}}\left|\sin \left[\frac{m_{\chi} n L}{2}\right] \sin \left[\frac{m_{\chi}(T-(n-1) L)}{2}\right] \sin \left[\frac{m_{\chi} T}{2}\right]\right|,
\end{equation}
We present projected sensitivities for the Atom Interferometer Observatory and Network (AION) in Fig.\,\ref{fig:atomic_constraint}, following the parameters in Refs.~\cite{Badurina:2019hst,Buchmueller:2023nll} and assuming solely photon coupling. Terrestrial configurations (AION-10, -100, -km) show significant sensitivity gains with increasing baseline, potentially probing $f_\chi\sim10^{30}$~GeV for masses $m_\chi\sim10^{-16}$~eV. However, terrestrial performance at low frequencies ($\omega_\chi\gtrsim0.1$~Hz) is saturated by gravity gradient noise (GGN). The proposed space-based detector AEDGE avoids this limitation, extending sensitivity to lower masses until GGN becomes dominant below $10^{-4}$~Hz. 

\section{Conclusions}
\label{sec:summary}
In this work, we have developed a systematic EFT framework for the dilaton, the pNGB arising from the spontaneous breaking of scale symmetry. Employing a scale-invariant regularization scheme, we derived the universal coupling of the dilaton to the trace anomaly, demonstrating consistency with the conformal compensator formalism while enabling a rigorous treatment of renormalization group evolution. Importantly, the running behavior of dilaton couplings constructed from the $\beta$-functions can be inferred from the conventional theory, as higher-loop differences are suppressed by inverse powers of the dilaton VEV and are thus negligible for the discussion in this work.

\begin{figure}
    \centering
    \includegraphics[width=\linewidth]{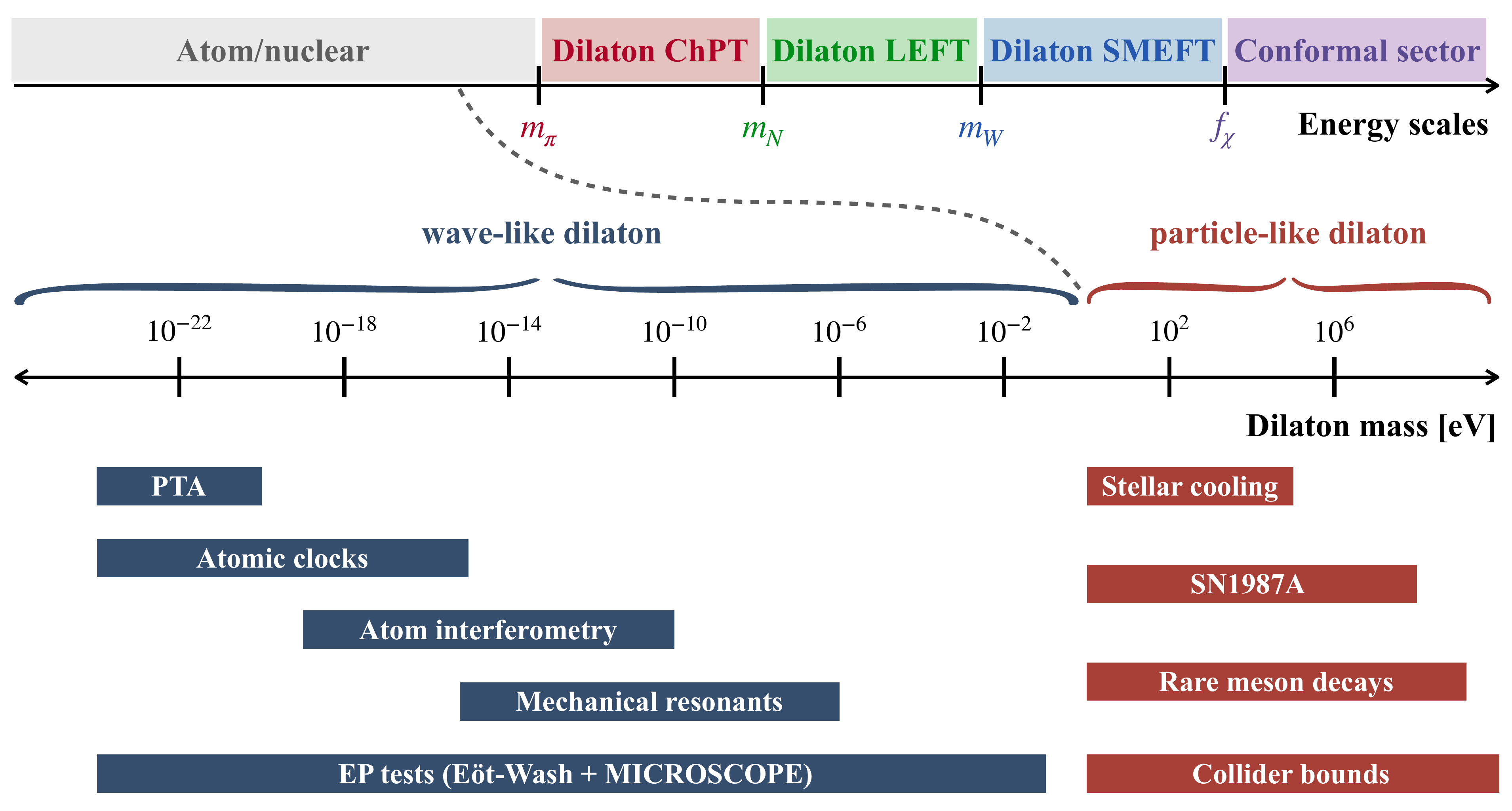}
    \caption{Top panel: Hierarchy of energy scales for the dilaton EFTs constructed in this work. Bottom panel: Summary of current and future laboratory experiments, cosmological observations and astrophysical measurements to set constraints on the dilaton parameter space. The mass spectrum is divided into two distinct regimes: wave-like and particle-like dilatons, demonstrating the complementary coverage of diverse experimental probes.}
    \label{fig:conclusion}
\end{figure}

Assuming a conformal sector in the UV, we determined the effective interactions of the dilaton with SM fields after the scale symmetry breaking. We extended the formalism to include higher-dimensional operators and constructed a consistent tower of EFTs across energy scales, as illustrated in the top panel of Fig.\,\ref{fig:conclusion}: from the dilaton-extended SMEFT near the electroweak scale, down to the LEFT below $m_W \sim 100$ GeV, and finally to $\chi$PT below $m_N \sim 1$ GeV. We explicitly provided a model-independent basis for dilaton operators in LEFT up to dimension 7 and identified the independent NLO operators in $\chi$PT, which can be matched to each other by the chiral symmetry.

We demonstrated the utility of this framework through a comprehensive phenomenological analysis across two distinct mass regimes, in which dilaton manifests as either conventional particle or wave-like particle. For MeV-scale dilatons behaving as conventional particles, we derived constraints from the HL-LHC ($pp\to j\chi$), Belle II ($B^+\to K^+\chi$) and NA62 ($K^+\to\pi^+\chi$). Additionally, we utilized the nucleon-dilaton coupling derived in $\chi$PT to evaluate its contribution to energy loss rate in the core of SN1987A. We found that laboratory experiments probe dilaton VEV $f_\chi$ of $\sim\mathcal{O}(10)\sim\mathcal{O}(1000)~$TeV for $m_\chi\in[0.1,200]~$MeV, while SN1987A excludes VEVs in the range $\sim\mathcal{O}(10^4)\sim\mathcal{O}(10^6)~$TeV. In the ultralight regime ($10^{-24}~\mathrm{eV}\lesssim m_\chi\lesssim10^{-10}~\mathrm{eV}$), where the dilaton behaves as wave-like DM, we showed projected sensitivities for atomic clocks (Yb$^+$/Al$^+$) and atom interferometers (AION stages and space-based AEDGE). We demonstrated that future experiments could reach sensitivity of $f_\chi$ up to $\mathcal{O}(10^{27})$~TeV with clocks and $\mathcal{O}(10^{32})$~TeV with space-based interferometry.

By systematically connecting high-energy collider searches with low-energy precision frontiers, this study establishes a robust foundation for identifying scale-invariant new physics. While the current analysis relies on leading-order matching, future efforts will address higher-order matching procedures to further refine the precision of these predictions.

Looking ahead, this unified framework paves the way for extended phenomenological studies across the full mass spectrum. We plan to broaden the scope of analysis as illustrated in the bottom panel of Fig.\,\ref{fig:conclusion}:
\begin{itemize}
\item Wave-like regime: Investigations of Pulsar Timing Arrays (PTA)~\cite{Kaplan:2022lmz} to probe the nanohertz frequency range, mechanical resonator experiments~\cite{Manley:2019vxy} to cover the kHz--GHz window, and Equivalence Principle (EP) tests~\cite{Berge:2017ovy,Hees:2018fpg} to constrain long-range scalar forces.

\item Particle-like regime: Refined astrophysical constraints from stellar cooling (the Sun, red giants, and white dwarfs)~\cite{Dev:2020jkh,Bottaro:2023gep} to bridge the sensitivity gap in the keV region.

\item Visible signatures: Analysis of additional rare meson decay channels, such as visible resonances in $B^+ \to K^+ \mu^+ \mu^-$~\cite{LHCb:2016awg}, which offer complementary sensitivity to the invisible channels studied here.

\end{itemize}
All the above experimental probes would provide a comprehensive search strategy on light dilaton particle.

\acknowledgments

We would like to thank Yandong Liu for helpful discussion. This work is supported by the National Key Research and Development Program of China Grant No. 2021YFA0718304, No. 2020YFC2201501, and the National Science Foundation of China under Grant Nos. 12347105, 12375099, 12235001, and 12447101. This work is also partly supported by Fundamental and Interdisciplinary Disciplines Breakthrough Plan of the Ministry of Education of China. The authors gratefully acknowledge the valuable discussions and insights provided by the members of the Collaboration on Precision Tests and New Physics (CPTNP).

\bibliographystyle{JHEP}
\bibliography{biblio.bib}

\providecommand{\href}[2]{#2}\begingroup\raggedright\begin{thebibliography}{100}

\bibitem{Kim:2008hd}
J.E.~Kim, G.~Carosi, Rev. Mod. Phys., {\bfseries 82}: 557--602 (2010).

\bibitem{DiLuzio:2020wdo}
L.~Di~Luzio, M.~Giannotti, E.~Nardi, \textit{et~al.}, Phys. Rept., {\bfseries
  870}: 1--117 (2020).

\bibitem{Peccei:2006as}
R.D.~Peccei, Lect. Notes Phys., {\bfseries 741}: 3--17 (2008).

\bibitem{Turner:1989vc}
M.S.~Turner, Phys. Rept., {\bfseries 197}: 67--97 (1990).

\bibitem{Brivio:2017ije}
I.~Brivio, M.B.~Gavela, L.~Merlo, \textit{et~al.}, Eur. Phys. J. C, {\bfseries
  77} (8): 572 (2017).

\bibitem{Galda:2021hbr}
A.M.~Galda, M.~Neubert, S.~Renner, JHEP, {\bfseries 06}: 135 (2021).

\bibitem{Bauer:2020jbp}
M.~Bauer, M.~Neubert, S.~Renner, \textit{et~al.}, JHEP, {\bfseries 04}: 063
  (2021).

\bibitem{Chala:2020wvs}
M.~Chala, G.~Guedes, M.~Ramos, \textit{et~al.}, Eur. Phys. J. C, {\bfseries 81}
  (2): 181 (2021).

\bibitem{Bonilla:2021ufe}
J.~Bonilla, I.~Brivio, M.B.~Gavela, \textit{et~al.}, JHEP, {\bfseries 11}: 168
  (2021).

\bibitem{Graham:2015ouw}
P.W.~Graham, I.G.~Irastorza, S.K.~Lamoreaux, \textit{et~al.}, Ann. Rev. Nucl.
  Part. Sci., {\bfseries 65}: 485--514 (2015).

\bibitem{Irastorza:2018dyq}
I.G.~Irastorza, J.~Redondo, Prog. Part. Nucl. Phys., {\bfseries 102}: 89--159
  (2018).

\bibitem{Ertas:2020xcc}
F.~Ertas, F.~Kahlhoefer, JHEP, {\bfseries 07}: 050 (2020).

\bibitem{Abel:2017rtm}
C.~Abel, N.J.~Ayres, G.~Ban, \textit{et~al.}, Phys. Rev. X, {\bfseries 7} (4):
  041034 (2017).

\bibitem{ADMX:2018gho}
N.~Du et~al (ADMX Collaboration), Phys. Rev. Lett., {\bfseries 120} (15):
  151301 (2018).

\bibitem{Salam:1969bwb}
A.~Salam, J.A.~Strathdee, Phys. Rev., {\bfseries 184}: 1760--1768 (1969).

\bibitem{Isham:1970gz}
C.J.~Isham, A.~Salam, J.A.~Strathdee, Phys. Lett. B, {\bfseries 31}: 300--302
  (1970).

\bibitem{Isham:1971dv}
C.J.~Isham, A.~Salam, J.A.~Strathdee, Annals Phys., {\bfseries 62}: 98--119
  (1971).

\bibitem{Ellis:1970yd}
J.R.~Ellis, Nucl. Phys. B, {\bfseries 22}: 478--492 (1970).

\bibitem{Ellis:1971sa}
J.R.~Ellis, Nucl. Phys. B, {\bfseries 26}: 536--546 (1971).

\bibitem{Appelquist:2016viq}
T.~Appelquist, R.C.~Brower, G.T.~Fleming, \textit{et~al.}, Phys. Rev. D,
  {\bfseries 93} (11): 114514 (2016).

\bibitem{LatKMI:2016xxi}
Y.~Aoki et~al (LatKMI Collaboration), Phys. Rev. D, {\bfseries 96} (1): 014508
  (2017).

\bibitem{LatKMI:2014xoh}
Y.~Aoki et~al (LatKMI Collaboration), Phys. Rev. D, {\bfseries 89}: 111502
  (2014).

\bibitem{Fodor:2016pls}
Z.~Fodor, K.~Holland, J.~Kuti, \textit{et~al.}, PoS, {\bfseries LATTICE2015}:
  219 (2016).

\bibitem{Fodor:2012ty}
Z.~Fodor, K.~Holland, J.~Kuti, \textit{et~al.}, Phys. Lett. B, {\bfseries 718}:
  657--666 (2012).

\bibitem{Appelquist:2017wcg}
T.~Appelquist, J.~Ingoldby, M.~Piai, JHEP, {\bfseries 07}: 035 (2017).

\bibitem{Appelquist:2017vyy}
T.~Appelquist, J.~Ingoldby, M.~Piai, JHEP, {\bfseries 03}: 039 (2018).

\bibitem{Appelquist:2019lgk}
T.~Appelquist, J.~Ingoldby, M.~Piai, Phys. Rev. D, {\bfseries 101} (7): 075025
  (2020).

\bibitem{Appelquist:2022mjb}
T.~Appelquist, J.~Ingoldby, M.~Piai, Universe, {\bfseries 9} (1): 10 (2023).

\bibitem{Appelquist:2025tol}
T.~Appelquist, J.~Ingoldby, M.~Piai,
  \href{https://arxiv.org/abs/2512.16863}{arXiv:2512.16863}.

\bibitem{Appelquist:2020bqj}
T.~Appelquist, J.~Ingoldby, M.~Piai, Phys. Rev. Lett., {\bfseries 126} (19):
  191804 (2021).

\bibitem{Appelquist:2022qgl}
T.~Appelquist, J.~Ingoldby, M.~Piai, Nucl. Phys. B, {\bfseries 983}: 115930
  (2022).

\bibitem{Appelquist:2024koa}
T.~Appelquist, J.~Ingoldby, M.~Piai, Phys. Rev. D, {\bfseries 110} (3): 035013
  (2024).

\bibitem{Goldberger:2007zk}
W.D.~Goldberger, B.~Grinstein, W.~Skiba, Phys. Rev. Lett., {\bfseries 100}:
  111802 (2008).

\bibitem{Appelquist:2010gy}
T.~Appelquist, Y.~Bai, Phys. Rev. D, {\bfseries 82}: 071701 (2010).

\bibitem{Vecchi:2010gj}
L.~Vecchi, Phys. Rev. D, {\bfseries 82}: 076009 (2010).

\bibitem{PhysRevD.87.115006}
Z.~Chacko, R.K.~Mishra, Phys. Rev. D, {\bfseries 87}: 115006 (2013).

\bibitem{Arvanitaki:2014faa}
A.~Arvanitaki, J.~Huang, K.~Van~Tilburg, Phys. Rev. D, {\bfseries 91} (1):
  015015 (2015).

\bibitem{VanTilburg:2015oza}
K.~Van~Tilburg, N.~Leefer, L.~Bougas, \textit{et~al.}, Phys. Rev. Lett.,
  {\bfseries 115} (1): 011802 (2015).

\bibitem{Hees:2016gop}
A.~Hees, J.~Gu{\'e}na, M.~Abgrall, \textit{et~al.}, Phys. Rev. Lett.,
  {\bfseries 117} (6): 061301 (2016).

\bibitem{Berge:2017ovy}
J.~Berg{\'e}, P.~Brax, G.~M{\'e}tris, \textit{et~al.}, Phys. Rev. Lett.,
  {\bfseries 120} (14): 141101 (2018).

\bibitem{Hees:2018fpg}
A.~Hees, O.~Minazzoli, E.~Savalle, \textit{et~al.}, Phys. Rev. D, {\bfseries
  98} (6): 064051 (2018).

\bibitem{Kennedy:2020bac}
C.J.~Kennedy, E.~Oelker, J.M.~Robinson, \textit{et~al.}, Phys. Rev. Lett.,
  {\bfseries 125} (20): 201302 (2020).

\bibitem{Filzinger:2023zrs}
M.~Filzinger, S.~D{\"o}rscher, R.~Lange, \textit{et~al.}, Phys. Rev. Lett.,
  {\bfseries 130} (25): 253001 (2023).

\bibitem{Geraci:2016fva}
A.A.~Geraci, A.~Derevianko, Phys. Rev. Lett., {\bfseries 117} (26): 261301
  (2016).

\bibitem{Arvanitaki:2016fyj}
A.~Arvanitaki, P.W.~Graham, J.M.~Hogan, \textit{et~al.}, Phys. Rev. D,
  {\bfseries 97} (7): 075020 (2018).

\bibitem{Badurina:2019hst}
L.~Badurina, E.~Bentine, D.~Blas, \textit{et~al.}, JCAP, {\bfseries 05}: 011
  (2020).

\bibitem{Buchmueller:2023nll}
O.~Buchmueller, J.~Ellis, U.~Schneider, Contemp. Phys., {\bfseries 64} (2):
  93--110 (2023).

\bibitem{Damour:2010rp}
T.~Damour, J.F.~Donoghue, Phys. Rev. D, {\bfseries 82}: 084033 (2010).

\bibitem{Nitti:2012ev}
F.~Nitti, F.~Piazza, Phys. Rev. D, {\bfseries 86}: 122002 (2012).

\bibitem{Low:2001bw}
I.~Low, A.V.~Manohar, Phys. Rev. Lett., {\bfseries 88}: 101602 (2002).

\bibitem{Callan:1970ze}
C.G.~Callan, Jr., S.R.~Coleman, R.~Jackiw, Annals Phys., {\bfseries 59}: 42--73
  (1970).

\bibitem{Tarrach:1981bi}
R.~Tarrach, Nucl. Phys. B, {\bfseries 196}: 45--61 (1982).

\bibitem{Tamarit:2013vda}
C.~Tamarit, JHEP, {\bfseries 12}: 098 (2013).

\bibitem{Ghilencea:2015mza}
D.M.~Ghilencea, Phys. Rev. D, {\bfseries 93} (10): 105006 (2016).

\bibitem{Collins:1984xc}
J.C.~Collins, \emph{{Renormalization}}, vol.~26 of \emph{Cambridge Monographs
  on Mathematical Physics} (Cambridge: Cambridge University Press, 2023).

\bibitem{Ghilencea:2016dsl}
D.M.~Ghilencea, Z.~Lalak, P.~Olszewski, Phys. Rev. D, {\bfseries 96} (5):
  055034 (2017).

\bibitem{Coleman:1973jx}
S.R.~Coleman, E.J.~Weinberg, Phys. Rev. D, {\bfseries 7}: 1888--1910 (1973).

\bibitem{Ferreira:2018itt}
P.G.~Ferreira, C.T.~Hill, G.G.~Ross, Phys. Rev. D, {\bfseries 98} (11): 116012
  (2018).

\bibitem{Oda:2021rcj}
I.~Oda, \href{https://arxiv.org/abs/2110.15408}{arXiv:2110.15408}.

\bibitem{Nogradi:2021zqw}
D.~Nogradi, B.~Ozsvath, SciPost Phys., {\bfseries 12} (5): 169 (2022).

\bibitem{Weinberg:1979sa}
S.~Weinberg, Phys. Rev. Lett., {\bfseries 43}: 1566--1570 (1979).

\bibitem{Abdullahi:2022jlv}
A.M.~Abdullahi, P.~Barham~Alz{\'a}s, B.~Brian, \textit{et~al.}, J. Phys. G,
  {\bfseries 50} (2): 020501 (2023).

\bibitem{Jenkins:2017jig}
E.E.~Jenkins, A.V.~Manohar, P.~Stoffer, JHEP, {\bfseries 03}: 016 (2018).

\bibitem{Jenkins:2017dyc}
E.E.~Jenkins, A.V.~Manohar, P.~Stoffer, JHEP, {\bfseries 01}: 084 (2018).

\bibitem{Liao:2020zyx}
Y.~Liao, X.-D.~Ma, Q.-Y.~Wang, JHEP, {\bfseries 08}: 162 (2020).

\bibitem{Li:2020tsi}
H.-L.~Li, Z.~Ren, M.-L.~Xiao, \textit{et~al.}, JHEP, {\bfseries 06}: 138
  (2021).

\bibitem{Murphy:2020cly}
C.W.~Murphy, JHEP, {\bfseries 04}: 101 (2021).

\bibitem{Goldstone:1961eq}
J.~Goldstone, Nuovo Cim., {\bfseries 19}: 154--164 (1961).

\bibitem{Goldstone:1962es}
J.~Goldstone, A.~Salam, S.~Weinberg, Phys. Rev., {\bfseries 127}: 965--970
  (1962).

\bibitem{Weinberg:1968de}
S.~Weinberg, Phys. Rev., {\bfseries 166}: 1568--1577 (1968).

\bibitem{Coleman:1969sm}
S.R.~Coleman, J.~Wess, B.~Zumino, Phys. Rev., {\bfseries 177}: 2239--2247
  (1969).

\bibitem{Callan:1969sn}
C.G.~Callan, Jr., S.R.~Coleman, J.~Wess, \textit{et~al.}, Phys. Rev.,
  {\bfseries 177}: 2247--2250 (1969).

\bibitem{Wess:1971yu}
J.~Wess, B.~Zumino, Phys. Lett. B, {\bfseries 37}: 95--97 (1971).

\bibitem{Witten:1983tw}
E.~Witten, Nucl. Phys. B, {\bfseries 223}: 422--432 (1983).

\bibitem{Gasser:1983yg}
J.~Gasser, H.~Leutwyler, Annals Phys., {\bfseries 158}: 142 (1984).

\bibitem{Gasser:1984gg}
J.~Gasser, H.~Leutwyler, Nucl. Phys. B, {\bfseries 250}: 465--516 (1985).

\bibitem{Fearing:1994ga}
H.W.~Fearing, S.~Scherer, Phys. Rev. D, {\bfseries 53}: 315--348 (1996).

\bibitem{Bijnens:1999hw}
J.~Bijnens, G.~Colangelo, G.~Ecker, Annals Phys., {\bfseries 280}: 100--139
  (2000).

\bibitem{Bijnens:1999sh}
J.~Bijnens, G.~Colangelo, G.~Ecker, JHEP, {\bfseries 02}: 020 (1999).

\bibitem{Ebertshauser:2001nj}
T.~Ebertshauser, H.W.~Fearing, S.~Scherer, Phys. Rev. D, {\bfseries 65}: 054033
  (2002).

\bibitem{Bijnens:2001bb}
J.~Bijnens, L.~Girlanda, P.~Talavera, Eur. Phys. J. C, {\bfseries 23}: 539--544
  (2002).

\bibitem{Li:2024ghg}
X.-H.~Li, H.~Sun, F.-J.~Tang, \textit{et~al.}, JHEP, {\bfseries 08}: 189
  (2024).

\bibitem{Jenkins:1990jv}
E.E.~Jenkins, A.V.~Manohar, Phys. Lett. B, {\bfseries 255}: 558--562 (1991).

\bibitem{Ecker:1995rk}
G.~Ecker, M.~Mojzis, Phys. Lett. B, {\bfseries 365}: 312--318 (1996).

\bibitem{Fettes:2000gb}
N.~Fettes, U.-G.~Meissner, M.~Mojzis, \textit{et~al.}, Annals Phys., {\bfseries
  283}: 273--302 (2000).

\bibitem{Kobach:2018pie}
A.~Kobach, S.~Pal, JHEP, {\bfseries 11}: 012 (2019).

\bibitem{Krause:1990xc}
A.~Krause, Helv. Phys. Acta, {\bfseries 63}: 3--70 (1990).

\bibitem{Fettes:1998ud}
N.~Fettes, U.-G.~Meissner, S.~Steininger, Nucl. Phys. A, {\bfseries 640}:
  199--234 (1998).

\bibitem{Frink:2004ic}
M.~Frink, U.-G.~Meissner, JHEP, {\bfseries 07}: 028 (2004).

\bibitem{Oller:2006yh}
J.A.~Oller, M.~Verbeni, J.~Prades, JHEP, {\bfseries 09}: 079 (2006).

\bibitem{Frink:2006hx}
M.~Frink, U.-G.~Meissner, Eur. Phys. J. A, {\bfseries 29}: 255--260 (2006).

\bibitem{Jiang:2016vax}
S.-Z.~Jiang, Q.-S.~Chen, Y.-R.~Liu, Phys. Rev. D, {\bfseries 95} (1): 014012
  (2017).

\bibitem{Song:2024fae}
C.-Q.~Song, H.~Sun, J.-H.~Yu, JHEP, {\bfseries 09}: 171 (2024).

\bibitem{Shamir:2021rav}
Y.~Shamir, M.~Golterman, PoS, {\bfseries LATTICE2021}: 372 (2022).

\bibitem{Golterman:2020tdq}
M.~Golterman, E.T.~Neil, Y.~Shamir, Phys. Rev. D, {\bfseries 102} (3): 034515
  (2020).

\bibitem{Song:2025snz}
C.-Q.~Song, H.~Sun, J.-H.~Yu,
  \href{https://arxiv.org/abs/2501.09787}{arXiv:2501.09787}.

\bibitem{Gavela:2019wzg}
M.B.~Gavela, R.~Houtz, P.~Quilez, \textit{et~al.}, Eur. Phys. J. C, {\bfseries
  79} (5): 369 (2019).

\bibitem{Izaguirre:2016dfi}
E.~Izaguirre, T.~Lin, B.~Shuve, Phys. Rev. Lett., {\bfseries 118} (11): 111802
  (2017).

\bibitem{Dev:2020jkh}
P.S.B.~Dev, R.N.~Mohapatra, Y.~Zhang, JCAP, {\bfseries 05}: 014 (2021).

\bibitem{Bottaro:2023gep}
S.~Bottaro, A.~Caputo, G.~Raffelt, \textit{et~al.}, JCAP, {\bfseries 07}: 071
  (2023).

\bibitem{Hook:2018dlk}
A.~Hook, PoS, {\bfseries TASI2018}: 004 (2019).

\bibitem{CMS:2011esc}
S.~Chatrchyan et~al (CMS Collaboration), Phys. Rev. Lett., {\bfseries 107}:
  201804 (2011).

\bibitem{CMS:2014jvv}
V.~Khachatryan et~al (CMS Collaboration), Eur. Phys. J. C, {\bfseries 75} (5):
  235 (2015).

\bibitem{CMS:2021far}
A.~Tumasyan et~al (CMS Collaboration), JHEP, {\bfseries 11}: 153 (2021).

\bibitem{Alloul:2013bka}
A.~Alloul, N.D.~Christensen, C.~Degrande, \textit{et~al.}, Comput. Phys.
  Commun., {\bfseries 185}: 2250--2300 (2014).

\bibitem{Alwall:2014hca}
J.~Alwall, R.~Frederix, S.~Frixione, \textit{et~al.}, JHEP, {\bfseries 07}: 079
  (2014).

\bibitem{Sjostrand:2014zea}
T.~Sj{\"o}strand, S.~Ask, J.R.~Christiansen, \textit{et~al.}, Comput. Phys.
  Commun., {\bfseries 191}: 159--177 (2015).

\bibitem{deFavereau:2013fsa}
J.~de~Favereau et~al (DELPHES 3 Collaboration), JHEP, {\bfseries 02}: 057
  (2014).

\bibitem{Cowan:2010js}
G.~Cowan, K.~Cranmer, E.~Gross, \textit{et~al.}, Eur. Phys. J. C, {\bfseries
  71}: 1554 (2011).

\bibitem{Gubernari:2018wyi}
N.~Gubernari, A.~Kokulu, D.~van Dyk, JHEP, {\bfseries 01}: 150 (2019).

\bibitem{Bharucha:2015bzk}
A.~Bharucha, D.M.~Straub, R.~Zwicky, JHEP, {\bfseries 08}: 098 (2016).

\bibitem{Parrott:2022rgu}
W.G.~Parrott et~al (HPQCD Collaboration), Phys. Rev. D, {\bfseries 107} (1):
  014510 (2023).

\bibitem{Grunwald:2023nli}
C.~Grunwald, G.~Hiller, K.~Kr\"oninger, \textit{et~al.}, JHEP, {\bfseries 11}:
  110 (2023).

\bibitem{Belle-II:2023esi}
I.~Adachi et~al (Belle-II Collaboration), Phys. Rev. D, {\bfseries 109} (11):
  112006 (2024).

\bibitem{Parrott:2022zte}
W.G.~Parrott et~al (HPQCD Collaboration), Phys. Rev. D, {\bfseries 107} (1):
  014511 (2023).

\bibitem{Abdughani:2023dlr}
M.~Abdughani, Y.~Reyimuaji, Phys. Rev. D, {\bfseries 110} (5): 055013 (2024).

\bibitem{Boyle:2010bh}
P.A.~Boyle et~al (RBC-UKQCD Collaboration), Eur. Phys. J. C, {\bfseries 69}:
  159--167 (2010).

\bibitem{Carrasco:2016kpy}
N.~Carrasco, P.~Lami, V.~Lubicz, \textit{et~al.}, Phys. Rev. D, {\bfseries 93}
  (11): 114512 (2016).

\bibitem{NA62:2021zjw}
E.~Cortina~Gil et~al (NA62 Collaboration), JHEP, {\bfseries 06}: 093 (2021).

\bibitem{Caputo:2022mah}
A.~Caputo, H.-T.~Janka, G.~Raffelt, \textit{et~al.}, Phys. Rev. Lett.,
  {\bfseries 128} (22): 221103 (2022).

\bibitem{Fiorillo:2025yzf}
D.F.G.~Fiorillo, T.~Pitik, E.~Vitagliano, Phys. Rev. Lett., {\bfseries 135}
  (7): 071005 (2025).

\bibitem{Diamond:2023cto}
M.~Diamond, D.F.G.~Fiorillo, G.~Marques-Tavares, \textit{et~al.}, Phys. Rev.
  Lett., {\bfseries 132} (10): 101004 (2024).

\bibitem{Diamond:2023scc}
M.~Diamond, D.F.G.~Fiorillo, G.~Marques-Tavares, \textit{et~al.}, Phys. Rev. D,
  {\bfseries 107} (10): 103029 (2023).

\bibitem{Dev:2020eam}
P.S.B.~Dev, R.N.~Mohapatra, Y.~Zhang, JCAP, {\bfseries 08}: 003 (2020).

\bibitem{Balaji:2022noj}
S.~Balaji, P.S.B.~Dev, J.~Silk, \textit{et~al.}, JCAP, {\bfseries 12}: 024
  (2022).

\bibitem{Kamiokande-II:1987idp}
K.~Hirata et~al (Kamiokande-II Collaboration), Phys. Rev. Lett., {\bfseries
  58}: 1490--1493 (1987).

\bibitem{Girmohanta:2023tdr}
S.~Girmohanta, Y.~Nakai, Y.~Shigekami, \textit{et~al.}, JHEP, {\bfseries 01}:
  153 (2024).

\bibitem{Krnjaic:2015mbs}
G.~Krnjaic, Phys. Rev. D, {\bfseries 94} (7): 073009 (2016).

\bibitem{Ishizuka:1989ts}
N.~Ishizuka, M.~Yoshimura, Prog. Theor. Phys., {\bfseries 84}: 233--250 (1990).

\bibitem{Kimball:2023vxk}
D.F.J.~Kimball, K.~van Bibber, eds., \emph{{The Search for Ultralight Bosonic
  Dark Matter}} (Springer, 2023).

\bibitem{Dzuba:1998au}
V.A.~Dzuba, V.V.~Flambaum, J.K.~Webb, Phys. Rev. A, {\bfseries 59}: 230--237
  (1999).

\bibitem{Antypas:2022asj}
D.~Antypas, A.~Banerjee, C.~Bartram, \textit{et~al.},
  \href{https://arxiv.org/abs/2203.14915}{arXiv:2203.14915}.

\bibitem{Safronova:2017xyt}
M.S.~Safronova, D.~Budker, D.~DeMille, \textit{et~al.}, Rev. Mod. Phys.,
  {\bfseries 90} (2): 025008 (2018).

\bibitem{Derevianko:2022lmn}
A.~Derevianko, E.~Figueroa, J.~Mart{\'\i}nez-Rinc{\'o}n, \textit{et~al.},
  {Quantum Networks for High Energy Physics}, in \emph{{Snowmass 2021}} (2022).

\bibitem{Cronin:2009zz}
A.D.~Cronin, J.~Schmiedmayer, D.E.~Pritchard, Rev. Mod. Phys., {\bfseries 81}:
  1051--1129 (2009).

\bibitem{Kaplan:2022lmz}
D.E.~Kaplan, A.~Mitridate, T.~Trickle, Phys. Rev. D, {\bfseries 106} (3):
  035032 (2022).

\bibitem{Manley:2019vxy}
J.~Manley, R.~Stump, D.~Wilson, \textit{et~al.}, Phys. Rev. Lett., {\bfseries
  124} (15): 151301 (2020).

\bibitem{LHCb:2016awg}
R.~Aaij et~al (LHCb Collaboration), Phys. Rev. D, {\bfseries 95} (7): 071101
  (2017).

\end{thebibliography}\endgroup

\end{document}